\documentclass[review]{elsarticle}

\usepackage{lineno,hyperref}
\usepackage{amsmath,graphicx}
\usepackage{mathtools}
\usepackage{cancel}
\usepackage{comment}
\usepackage{capt-of}
\usepackage{color}

\modulolinenumbers[500]

\journal{Nuclear Physics A}

\begin{document}

\begin{frontmatter}

\title{Constraining heavy quark energy loss using $B$ and $D$ meson measurements in heavy ion collision at RHIC and LHC energies}

\author[ad1]{Kapil Saraswat}
\author[ad2,ad3]{Prashant Shukla\corref{ca}}
  \cortext[ca]{Corresponding author}
  \ead{pshukla@barc.gov.in}
\author[ad1]{Venktesh Singh}

\address[ad1]{Department of Physics, Banaras Hindu University, Varanasi 221005, India.}
\address[ad2] {Nuclear Physics Division, Bhabha Atomic Research Centre, Mumbai 400085, India.}
\address[ad3]{Homi Bhabha National Institute, Anushakti Nagar, Mumbai 400094, India.}


%


\begin{abstract}
  In this work, we calculate energy loss of heavy quark (charm and bottom) 
due to elastic collisions and gluon radiation 
in hot/dense medium. The collisional energy loss has 
been obtained using QCD calculations. The radiative energy loss 
is calculated using reaction operator formalism and generalized 
dead cone approach. We rederive the energy loss expression using same 
assumptions as generalized dead cone approach but obtain slightly different 
results. We also improve the model employed to calculate path length and the system 
evolution. The nuclear modification factors $R_{AA}$ including shadowing and 
energy loss are evaluated for $B$ and $D$ mesons and are compared with the 
measurements in PbPb collision at $\sqrt{s_{NN}}$ = 2.76 TeV and with the D meson
and Heavy flavour (HF) electrons measurements in AuAu collision at 
$\sqrt{s_{NN}}$ = 200 GeV. 
 The radiative energy loss calculated by reaction operator formalism added 
with collisional energy loss describes the RHIC HF electron suppression 
in high $p_{T}$ range. It also describes the LHC measurement of $B$ meson suppression 
but overestimates the suppression of $D$ meson. The radiative energy loss from 
generalized dead cone approach describes the charm suppression at both 
RHIC as well as LHC energies and requires energy loss due to collisions to be 
added in order to describe the bottom suppression at LHC. \\
\end{abstract}


\end{frontmatter}

\linenumbers

\section{Introduction}
The heavy ion collisions at ultra relativistic energy create matter with 
high energy density required to form Quark Gluon Plasma (QGP).
 Relativistic Heavy Ion Collider (RHIC) and Large Hadron Collider (LHC) 
are designed to create and explore QGP. Many measurements at RHIC and LHC
 already point to the formation of QGP \cite{quarkmatter2014}. The heavy quarks 
(charm and bottom) are produced in hard partonic interactions in heavy ion 
collisions and their initial momentum distribution can be calculated 
from pQCD \cite{kumar2010}.  While traversing the hot/dense medium formed in the 
collisions, these quarks loose energy either due to the elastic collisions with 
the plasma constituents or by radiating a gluon or both. There are several 
formulations to calculate collisional 
\cite{bjorken1982, braaten1991,thoma1991, peshier2006, peigne2008} 
as well as radiative energy loss 
\cite{armesto2004, Armesto:2004vz, glv, dg}. For a review of 
many of these formalism see Ref.~\cite{Jamil2010}. At high parton energies, the radiative 
energy loss becomes much larger than the collisional energy loss but at lower energies, 
these two processes can contribute 
equally with the collisional energy loss being the dominant for small values 
of the parton energy \cite{ducati2007}. 

 There are many heavy quark energy loss models, each having  specific set of 
simplifications/assumptions. 
 The model by Gyulassy, Levai and Vitev (GLV) \cite{gyulassy2000,gyulassy2001}
 is based on a systematic expansion of the energy loss in terms of the 
number of scatterings and generally leading order term is included in the 
current calculations. The medium is characterized by two parameters,
the density of scattering centers or mean free path and Debye screening 
mass. Such an approach includes the interference between vacuum and 
medium induced radiation.  This formalism was then extended to obtain 
the energy loss for heavy quarks in Ref. \cite{dg} and was simplified for the 
first order of opacity expansion in Ref.~\cite{wicks}\\

  In this work, we calculate the radiative energy loss of heavy quarks 
(both charm or bottom quark) using reaction operator formalism 
 DGLV (Djordjevic, Gyulassy, Levai and Vitev) \cite{glv,dg,wicks}
and using generalized dead cone approach AJMS (Abir, Jamil, Mustafa and 
Srivastava) \cite{rumd}. 
  We rederive the energy loss expression using same 
assumptions as generalized dead cone approach but obtain slightly different 
results. We also improve the model employed to calculate path length and the 
system evolution.
 The collisional energy loss has been calculated using Peigne and Peshier 
formalism \cite{peigne2008}.
  The nuclear modification factors $R_{AA}$ including shadowing and energy loss 
are evaluated for $B$ and $D$ mesons and are compared with the measurements in 
PbPb collision at $\sqrt{s_{NN}}$ = 2.76 TeV and with the HF electron measurement 
of PHENIX and $D$ meson measurements of STAR in AuAu collision at $\sqrt{s_{NN}}$ = 200 GeV.

\section{Heavy Quark Production by Hard Processes}

  The production cross sections of $c\bar c$ and $b\bar b$ pairs
 are calculated to NLO in pQCD using the CT10 parton densities \cite{Lai:2010vv}. 
 We use the same set of parameters as that of Ref.~ \cite{Nelson:2012bc}
which are obtained by fitting the energy dependence 
of open heavy flavor production to the measured total cross sections.
The charm quark mass and scale parameters used are $m_c = 1.27$~GeV, 
$\mu_F/m_{T\,c} = 2.10 $, 
and $\mu_R/m_{T\, c} = 1.60$~\cite{Nelson:2012bc}. 
The bottom quark mass and scale parameters are $m_b = 4.65$ GeV,
$\mu_F/m_{T\, b} = 1.40$, and $\mu_R/m_{T\, b} = 1.10$.
 Here $\mu_{F}$ is the factorization scale,~ $\mu_{R}$ 
is the renormalization scale and $m_{T} = \sqrt{M^{2} + p^{2}_{T}}$.
The central EPS09 NLO parameter set~\cite{eskola2009} is used to 
calculate the modifications of the parton distribution functions (nPDF) in 
heavy ion collisions, referred as shadowing effects.
  
For the fragmentation of heavy quarks into mesons, Peterson fragmentation function
is used which is given as follows \cite{Peterson:1982ak}
\begin{equation}
D_{Q}(z) = \frac{N}{z[1-(1/z)-\epsilon_{Q}/(1-z)]^{2}}.
\end{equation}
 Here $z=p_T^D/p_T^c$ and $N$ is normalization constant which is fixed by summing 
over all hadrons 
containing heavy quark,
\begin{equation}
\sum \int dz D_{Q}(z) = 1.
\end{equation}
 We take $\epsilon_{c}$ = 0.016 and $\epsilon_{b}$ = 0.0012. The schemes of $D$ meson to 
electron decay
(BR = 10.3 \%) and $B$ meson to $J/\psi$ decay (BR = 1.1 \%) are obtained by Pythia 
simulations \cite{Sjostrand:2006za}.\\

\begin{figure}[htp]
\centering
\includegraphics[width=0.65\linewidth]{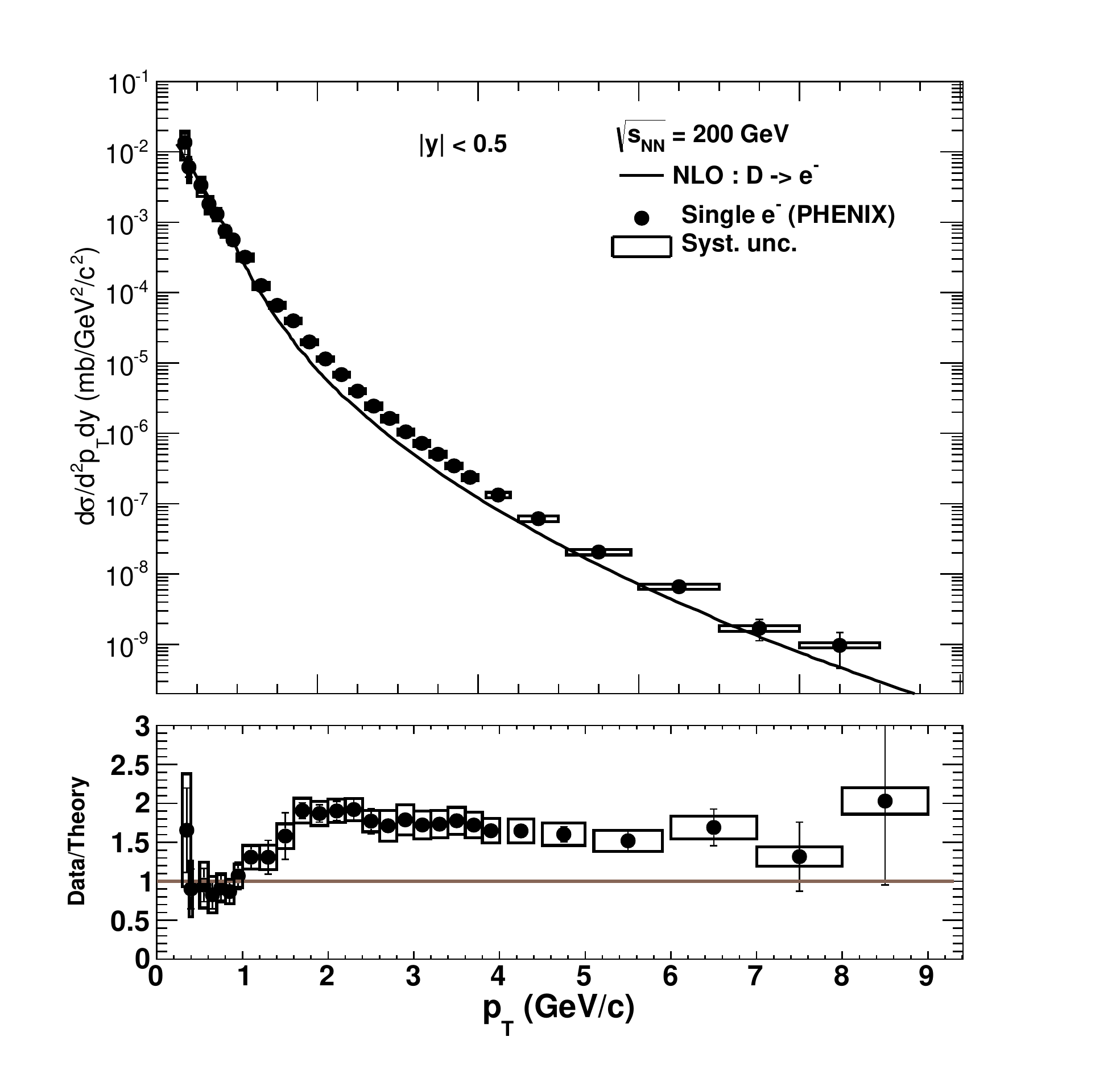}
\caption{(color online): The NLO calculations of differential cross section of 
single electrons from D mesons as a function of the transverse momentum $p_{T}$ 
in pp collision at $\sqrt{s}=200$ GeV. The data is from PHENIX measurements
of single electrons from heavy flavour~\cite{adare}.}
\label{figure1charm200ptdist}
\end{figure}

  Figure~$\ref{figure1charm200ptdist}$ shows the 
 NLO calculations of differential cross section of 
single electrons from $D$ mesons as a function of the transverse momentum $p_{T}$ 
in pp collision at $\sqrt{s}=200$ GeV compared with the PHENIX measurements
of single electrons from heavy flavour \cite{adare}. As shown in the 
Data/Theory panel the agreement between the data
and the calculations is not very good but since at $p_T$ above 2 GeV/$c$ the 
shapes of the calculations and the data are same, it does not affect the 
$R_{AA}$ calculations due to energy loss.
  
\begin{figure}[htp]
\centering
\includegraphics[width=0.60\linewidth]{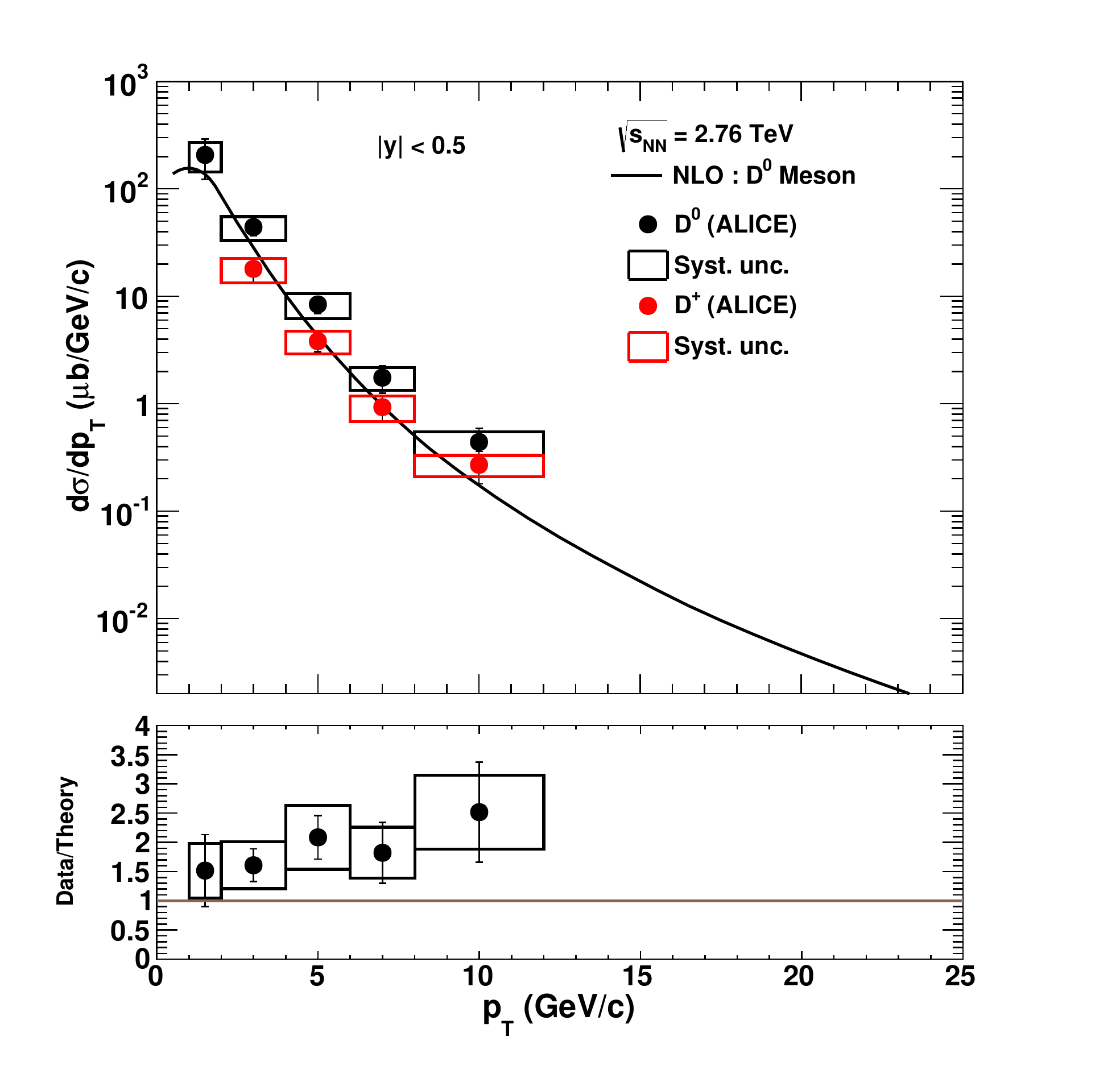}
\caption{(color online): 
 The NLO calculations of differential cross section of $D^{0}$ mesons as a 
function of the transverse momentum $p_{T}$ in pp collision at $\sqrt{s}=2.76$ TeV. 
The data is from ALICE measurements of $D^{0}$ and $D^{+}$ mesons \cite{Abelev:2012vra}.}
\label{figure2charm276ptdist}
\end{figure}

  Figure~$\ref{figure2charm276ptdist}$ shows
the NLO calculations of differential cross section of $D^{0}$ mesons as a 
function of the transverse momentum $p_{T}$ in pp collision at $\sqrt{s}=2.76$ TeV
compared with ALICE measurements of $D^{0}$ and $D^{+}$ mesons \cite{Abelev:2012vra}.
  Here also Data/Theory panel shows the agreement between the data
and the calculations is not very good but since at $p_T$ above 2 GeV/$c$ the 
shapes of the calculations and the data are same, it does not affect the 
$R_{AA}$ calculations due to energy loss.

\begin{figure}[htp]
\centering
\includegraphics[width=0.60\linewidth]{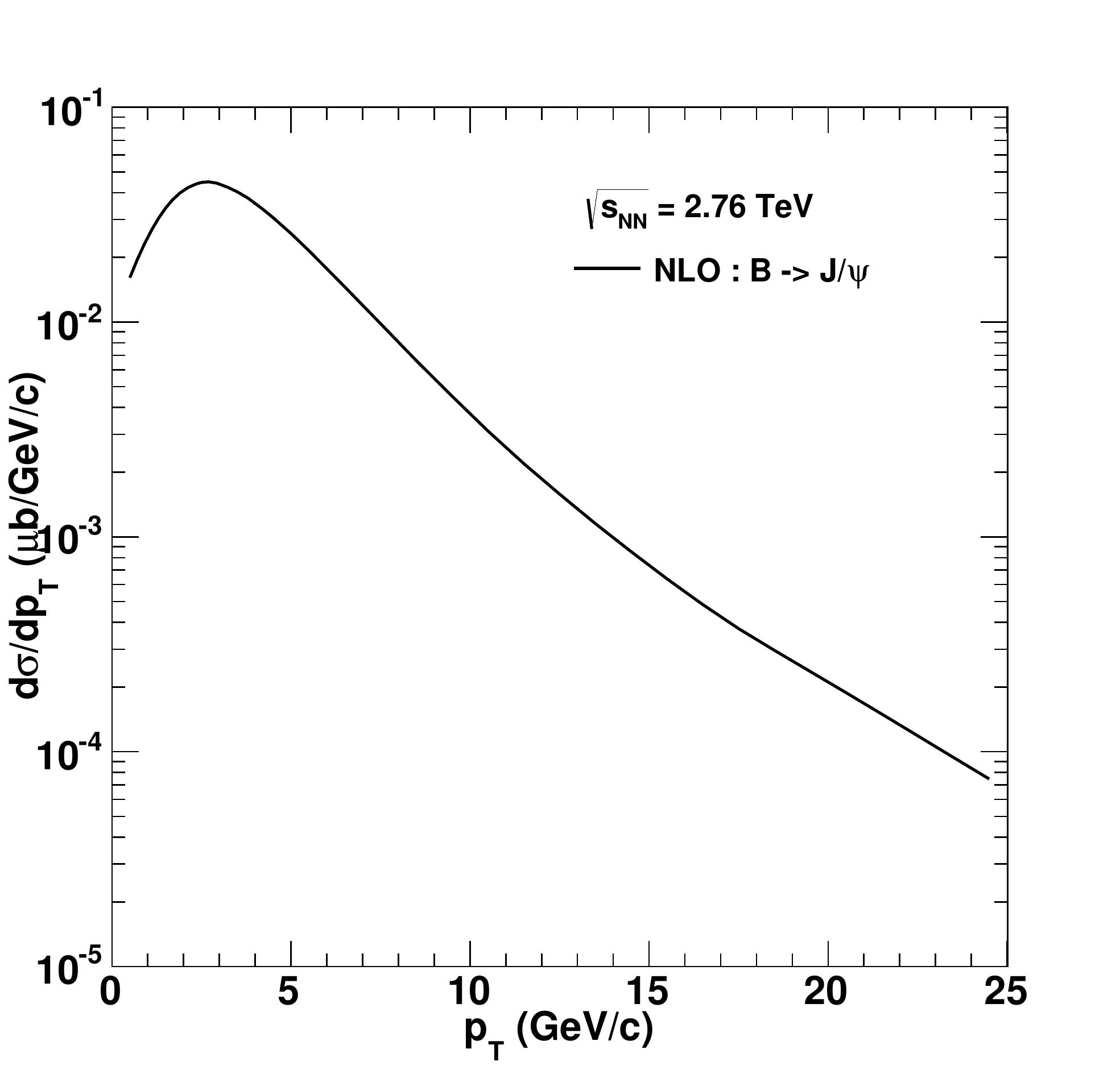}
\caption{(color online): 
 The NLO calculations of differential cross section of inclusive $J/\psi$ coming from 
B mesons as a function of the transverse momentum $p_{T}$ in pp collision at 
$\sqrt{s}=2.76$ TeV. }
\label{figure3bottom276ptdist}
\end{figure}

    Figure~$\ref{figure3bottom276ptdist}$ shows 
 the NLO calculations of differential cross section of inclusive J/$\psi$ coming from 
$B$ mesons as a function of the transverse momentum $p_{T}$ in pp collision at 
$\sqrt{s}=2.76$ TeV.

\section{\bf Collisional Energy Loss}
 The QCD calculation of the rate of energy loss of heavy quark per unit distance 
($dE/dx$) in QGP is given by Braaten and Thoma \cite{braaten1991}. 
 Their formalism is an extension of QED calculation of $dE/dx$ for a muon 
\cite{peshier2006} which assumes that the momentum exchange $q \ll E$. 
  Such an assumption is not valid in the domain when the energy of the heavy quark 
$E \gg M^{2}/T$, where $M$ is the mass of the heavy quark. 
  Peigne and Peshier (PP) \cite{peigne2008} extended this calculation 
which is valid in the domain $E \gg M^{2}/T$ to give the expression 
for $dE/dx$ as

\begin{equation}
\label{ppcollisionalloss}
\frac{dE}{dx} = \frac{4 \pi \alpha^{2}_{s} T^{2}}{3}~\Bigg[\Big(1+\frac{N_{f}}{6}\Big)
\log\Big(\frac{E T}{\mu^{2}_{g}}\Big) + \frac{2}{9}\log\frac{E T}{M^{2}} + 
c(N_{f})\Bigg]~.
\end{equation}
Here $\mu_{g}=\sqrt{4\pi \alpha_{s} T^{2}\Big(1+N_{f}/6\Big)}$ is the Debye screening 
mass and $c(N_{f}) \approx 0.146 N_{f} + 0.05$~. 
 $\alpha_{s}(=0.3)$ is the fine structure splitting 
constant for strong interaction and $N_f$ is the number of quark flavours.

\section{Radiative Energy Loss}
\subsection{\bf DGLV Formalism}
\noindent
 The energy loss of fast partons is dominated by radiation of gluons.
The reaction operator formalism is used in Ref.~\cite{glv} to obtain
the energy loss due to gluon radiation for light quark jets. Analytical 
expression is obtained for energy loss in powers of gluon opacity 
($L/\lambda$) where $\lambda$ is the mean free path of the quark and 
$L$ is the path length traversed in the medium. This formalism was 
then extended to obtain the energy loss for heavy quarks in Ref. \cite{dg} 
and was simplified for the first order of opacity expansion in 
Ref.~\cite{wicks}. The expression of the average radiative energy loss of 
heavy quark is given in appendix {\bf A}.

\subsection{\bf Generalized Dead Cone Approach (AJMS)}

The rate of radiative energy loss of a heavy quark with energy $E$ due to the 
inelastic scattering with the medium is calculated as
\begin{equation}
\label{averageE}
\frac{dE}{dx} = \frac{<\omega>}{\lambda},
\end{equation}
\noindent
where $<\omega>$ is the mean energy of the emitted gluons.

\noindent 
The probability of gluon emission off a heavy quark is written as ~\cite{rumd} 
\begin{equation}
\frac{d\eta_{g}}{d\eta~d\omega}=\frac{2C_{A}\alpha_{s}}{\pi}~
\frac{\mathcal D}{\omega}~~,
\end{equation}
where $C_{A}$(=3) is the Casimir operator in QCD and
$\omega$ is related to the transverse momentum of the emitted gluons
$k_{\perp}$ by the relation $k_{\perp}=\omega~\sin \theta$, where $\theta$ 
is the emission angle. $\mathcal D$ is the generalised dead cone which 
can be written 
as ~\cite{mustafa} 
\begin{equation}
\label{cone}
\mathcal D = \Bigg(1+\frac{M^{2}}{s}~e^{2\eta} \Bigg)^{-2}  ~~,~
\eta=-\ln\tan\Big(\frac{\theta}{2}\Big)~~.
\end{equation}
Here $s$ is mandelstam variable which is related to the energy $E$ and 
mass $M$ of heavy quark by the relation, $s=2E^2+2E\sqrt{E^2-M^2}-M^2$.

The mean energy of the emitted gluon can be written as ~\cite{rumd}
\begin{equation}
\label{omegaav}
<\omega>= \frac{\int~\frac{d\eta_{g}}{d\eta d\omega}~\omega~d\eta~d\omega}
{\int~\frac{d\eta_{g}}{d\eta d\omega}~d\eta~d\omega}  ~=~ 
\frac{\int d\omega~\int~\mathcal D~d\eta}{\int \frac{1}{\omega}~
d\omega~\int \mathcal D~ d\eta}~~.
\end{equation}

\noindent
The mean free path length $\lambda$ is calculated as ~\cite{rumd}
\begin{eqnarray}
\label{lambda}
\frac{1}{\lambda} &=& \rho_{q}~\sigma_{Qq(\bar q)\rightarrow Qq(\bar q)g} + 
\rho_{g}~\sigma_{Qg\rightarrow Qgg}~,  \\
 &=&   (\rho_{q}+\frac{9}{4}~\rho_{g}) ~\sigma_{2 \rightarrow 3}, \\
 &=& \rho_{QGP}~\sigma_{2 \rightarrow 3}~.
\end{eqnarray}

\noindent
The total cross section of the process $2\rightarrow3$ is calculated as
~\cite{bdmtw}
\begin{equation}
\label{xsection}
\sigma_{2\rightarrow 3} = 4~C_{A}~\alpha^{3}_{s}~\int~\frac{1}{(q^{2}_{\perp})^{2}}~
dq^{2}_{\perp}~\int~\frac{1}{\omega}~d\omega~\int~\mathcal D~d\eta~~.
\end{equation}
\noindent
Here $q_{\perp}$ is the transverse momentum of the exchanged gluon.
Using Eqs. ~$(\ref{averageE})$,~$(\ref{omegaav})$,~$(\ref{lambda})$ and 
$(\ref{xsection})$ and assigning the limits of the 
variables of $q^{2}$,~$\omega$ and $\eta$ we get
\begin{equation}
\label{energyperunitdistance}
\frac{dE}{dx}=24~\alpha^{3}_{s}~\rho_{QGP}~\int^{q^{2}_{\perp}|_{max}}_{q^{2}_{\perp}|_{min}}~
\frac{1}{(q^{2}_{\perp})^{2}}~dq^{2}_{\perp}~\int^{\omega_{max}}_{\omega_{min}}~d\omega~
~\int^{\eta_{max}}_{\eta_{min}}~\mathcal D~d\eta~~.
\end{equation}

\noindent
Here we have put $C_{A}$=3 and factor of $2$ is used to cover both upper and lower 
hemispheres of $\eta$.

\noindent
Equation $(\ref{energyperunitdistance})$ is solved to get following result 
(see details in appendix {\bf B}) which we call corrected AJMS result
\begin{equation}
\frac{dE}{dx}=24~\alpha^{3}_{s}~\rho_{QGP}~\frac{1}{\mu_{g}}~\Big(1-\beta_{1}\Big)
~\Bigg(\sqrt{\frac{1}{(1-\beta_{1})}~\log\Big(\frac{1}{\beta_{1}}\Big)}-1 \Bigg)
~\mathcal F(\delta)~~.
\end{equation}
Here
\begin{equation}
\label{integral3}
\mathcal F(\delta)=2\delta-\frac{1}{2}~\log\Bigg(
\frac{1+\frac{M^2}{s}~e^{2\delta}}{1+\frac{M^2}{s}~e^{-2\delta}}\Bigg)-
\Bigg(\frac{\frac{M^2}{s}~\sinh(2\delta)}
{1+2~\frac{M^2}{s}\cosh(2\delta)+\frac{M^4}{s^{2}}}\Bigg)~~.
\end{equation}
and 
\begin{equation}
\delta=\frac{1}{2}~\log\Bigg[\frac{1}{(1-\beta_{1})}~\log\Big(\frac{1}
{\beta_{1}}\Big)~\Bigg(1+\sqrt{1-\frac{(1-\beta_{1})}{\log(\frac{1}{\beta_{1}})}} 
\Bigg)^{2} \Bigg].
\end{equation}

The above results differs with the original AJMS calculation~\cite{rumd}
where the $\mathcal F(\delta)$ term is given by 

\begin{eqnarray}
\cal F(\delta) &=& ~2\delta-\frac{1}{2}\log\Bigg(\frac{1+\frac{M^2}{s}~e^{2\delta}}
{1+\frac{M^2}{s}e^{-2\delta}}\Bigg)- \frac{\frac{M^2}{s}\cosh(\delta)}
{1+2~\frac{M^2}{s}~\cosh(\delta) + \frac{M^4}{s^{2}}},  \nonumber \\
\delta &=& \frac{1}{2}\log\Bigg[\frac{1}{(1-\beta_{1})}~~\log\Big(\frac{1}
{\beta_{1}}\Big)~
\Bigg(1+\sqrt{1-\frac{(1-\beta_{1})^{\frac{1}{2}}}{[\log(\frac{1}
{\beta_{1}})]^{\frac{1}{2}}}} \Bigg)^{2}\Bigg]~.
\end{eqnarray}

\section{\bf Model For QGP Evolution}
  To estimate the energy loss suffered by the heavy quark, it is crucial to 
calculate its path length which it travel in the medium. 
 Let us assume that the heavy quark is produced at a point ($r$,~$\phi$) in heavy ion
collision, moves at an angle $\phi$ with respect to $\hat{\rm {r}}$ in the 
transverse plane. If $R$ is the radius of the colliding nuclei, then the distance $d$ 
covered by the heavy quark in the plasma is given ~\cite{muller} by
\begin{eqnarray}
d(\phi,\, { r})= \sqrt{{ R}^2\,-\,{ r}^2\sin^2 \phi}\,-\,{ r}\cos\phi~~.
\label{quarlklength}
\end{eqnarray}
\noindent
The average distance travelled by the heavy quark in the
plasma 
\begin{equation}
L=\frac{\int^{R}_{0}~\int^{2\pi}_{0}~d(\phi,r)~\rho(|\vec r|)~
\rho(|\vec r-\vec b|)~r~dr~d\phi}
{\int^{R}_{0}~\int^{2\pi}_{0}~\rho(|\vec r|)~\rho(|\vec r-\vec b|)~r~dr~d\phi}~~.
\end{equation}
 Here $\rho(|\vec r|)$ is the density of nucleus assumed to be a sharp sphere 
with radius $R=1.1~A^{1/3}$. If the velocity of the heavy quark is $v_{T}=p_{T}/m_{T}$, 
where $m_{T}$ is the transverse mass, the effective path length 
$L_{eff}$ is obtained as

\begin{equation}
L_{eff} = {\rm min} \Big[L ,~ v_{T}~ \times ~\tau_{f} \Big].
\end{equation}

  The evolution of the system for each centrality bin is governed by an 
isentropic cylindrical expansion with prescription given in 
Ref.~\cite{ZhaoRapp2011}.
 The entropy conservation condition $s(T)\,V(\tau)= s(T_0)\,V(\tau_0)$ 
and equation of state obtained by Lattice QCD along with hadronic resonance 
are used to obtain temperature as a function of proper 
time \cite{vineet}. 
 The transverse size $R$ for a given centrality with number of participant $N_{part}$ 
is obtained as  $R(N_{\rm part}) = R_{A} ~ \sqrt{2~A/N_{\rm part}}$,~ where $R_{A}$ is radius 
of the nucleus. The initial entropy density $s(\tau_0)$ is 
\begin{eqnarray}
s(\tau_0)  = {a_{\rm m} \over V(\tau_0)}   \left(\frac{dN}{d\eta} \right) . 
\end{eqnarray}  
 Here $a_m=5$ is a constant which relates the total entropy with the 
multiplicity \cite{Shuryak:1992wc}. The initial volume 
$V(\tau_0) = \pi \left[R(N_{\rm part})\right]^2 \tau_0$ and measured values of $dN/d\eta$~
for LHC ~\cite{Aamodt:2010cz} and for RHIC~ \cite{Back:2002uc} are used for a given 
centrality.\\

\begin{center}
\captionof{table}{Parameters of QGP evolution model}\label{parametertable}
\begin{tabular}{| c || c | c | c | c | c |}
\hline
Model               & Present  & Present & Present  & Model            & Model \\    
                    &          &         &          &\cite{wicks,rumd} & \cite{wicks,rumd} \\ \hline \hline    
Centrality $(\%)$   & 0-10     &  0-100  &  0-20    & 0-10             & 0-20 \\ \hline 
Experiment          & RHIC     &  LHC    &  LHC     & RHIC             & LHC  \\ \hline \hline
$b$ (fm)            & 3.26     &  9.72  &  4.70    & 0.0              & 0.0 \\ \hline
$N_{\rm part}$        & 329      &  113    &  308     &  -               & -  \\ \hline
$dN/d\eta$         & 623      &  360    &  1206    & -                 & - \\ \hline
$L$(fm)            & 5.63     &  4.3    &  5.62    & 5.78              & 6.14 \\ \hline
$\tau_{0}$(fm/c)    & 0.6     &  0.3    &  0.3      & 0.2               & 0.2 \\ \hline
$\tau_{f}$(fm/c)    & 3.0     &  6.0    &  6.0      & 2.63              & 5.90 \\ \hline
$T_{0}$(GeV)        & 0.303   &  0.450  &  0.481    & 0.400             & 0.525 \\ \hline
\end{tabular}
\end{center}

 We calculate the energy loss which is then averaged over the 
temperature evolution.
Various parameters used and calculated in our model for different centralities 
such as number of participant $N_{\rm part}$, measured $dN/d\eta$,
calculated average path length $(L)$, initial time $(\tau_{0})$, QGP life 
time $(\tau_{f})$ and initial temperature $(T_{0})$ are given in 
Table~\ref{parametertable}. 
The parameters used in the earlier model \cite{wicks,rumd} are also given.

\section{\bf Results and Discussions}
\begin{figure}[htp]
\centering
\includegraphics[width=0.60\linewidth]{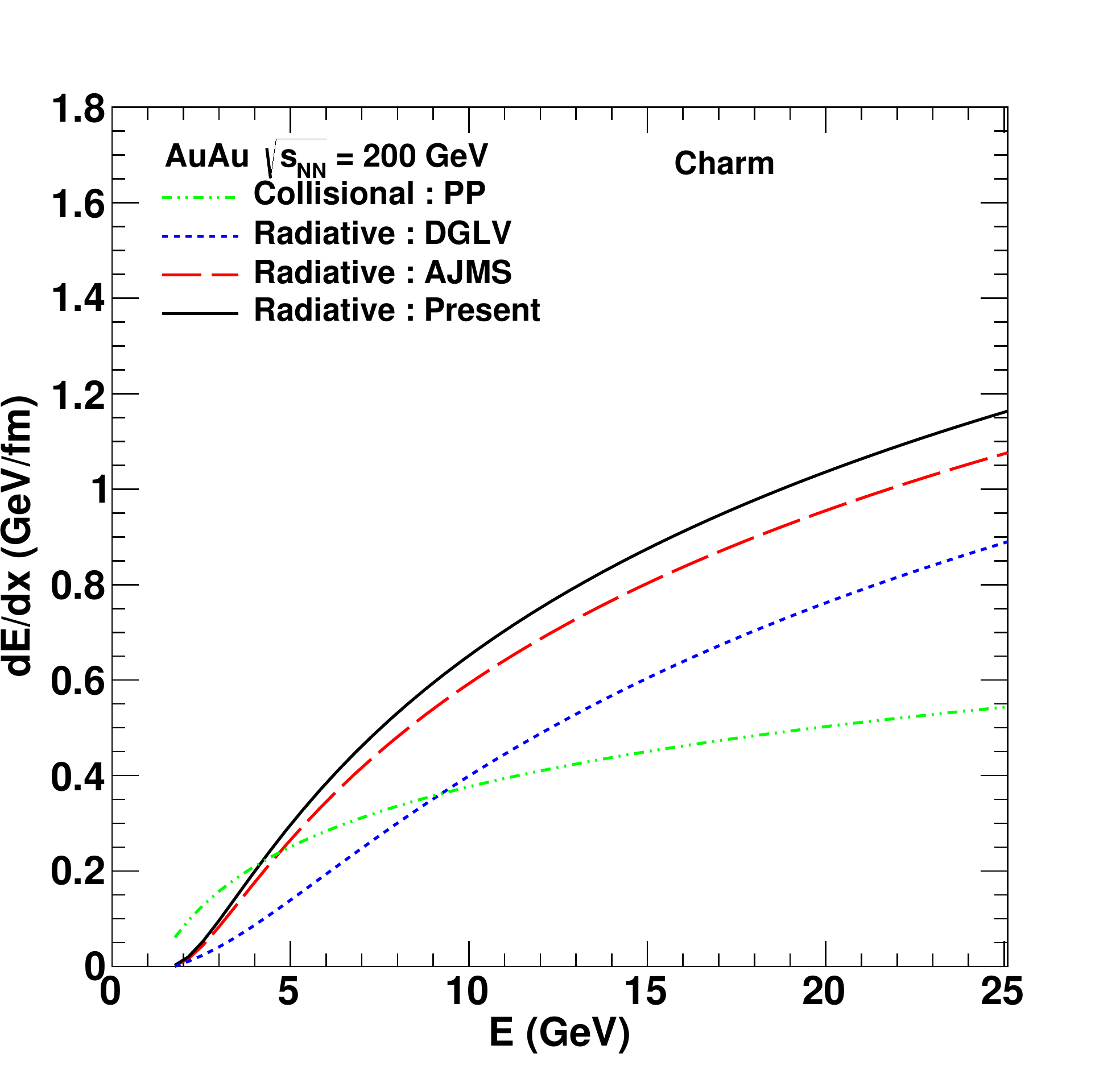}
\caption{(color online): The energy loss $dE/dx$ as a function of energy
 of charm quark obtained using PP, DGLV, AJMS and corrected 
AJMS (Present) calculations for AuAu collision at $\sqrt{s_{NN}}$=200 GeV.}
\label{figure4charm200energyloss}
\end{figure}

\begin{figure}[htp]
\centering
\includegraphics[width=0.60\linewidth]{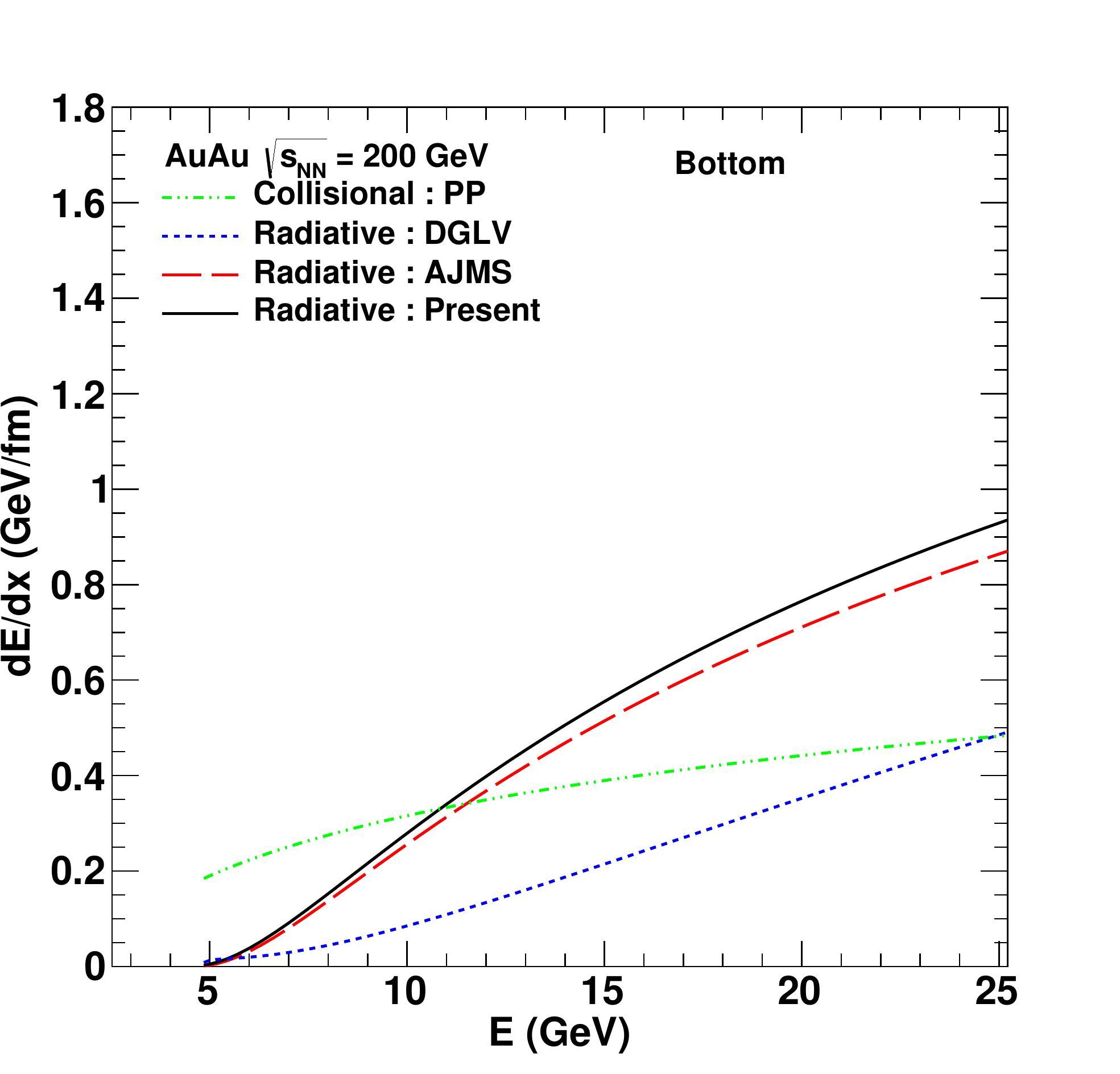} 
\caption{(color online): The energy loss $dE/dx$ as a function of energy
 of bottom quark obtained using PP, DGLV, AJMS and corrected 
AJMS (Present) calculations for AuAu collision at $\sqrt{s_{NN}}$=200 GeV.}
\label{figure5bottom200energyloss}
\end{figure}

\begin{figure}[htp]
\centering
\includegraphics[width=0.60\linewidth]{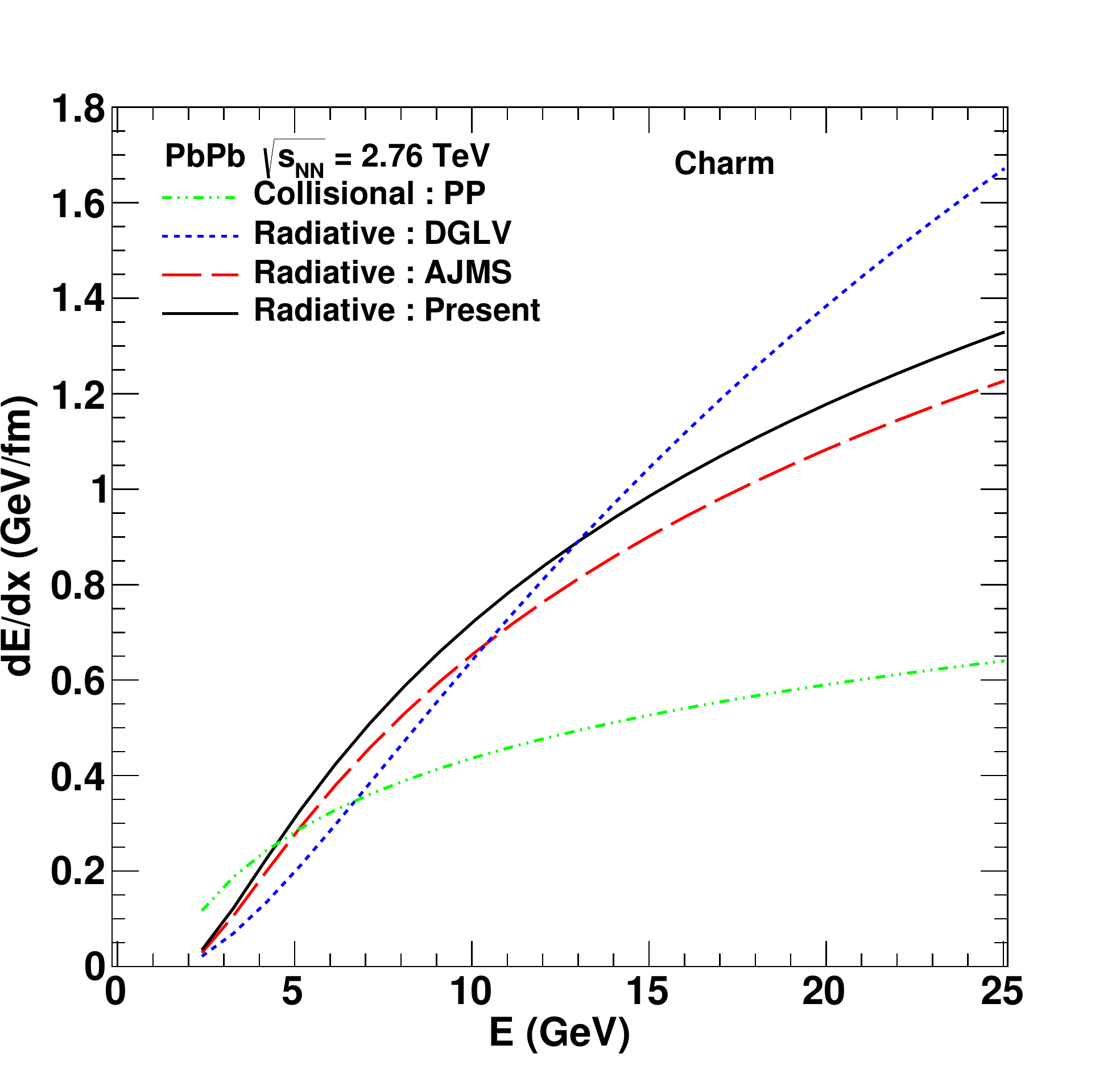} 
\caption{(color online): The energy loss $dE/dx$ as a function of energy
 of charm quark obtained using PP, DGLV, AJMS and corrected 
AJMS (Present) calculations for PbPb collision at $\sqrt{s_{NN}}$=2.76 TeV.}
\label{figure6charm276energyloss}
\end{figure}

\begin{figure}[htp]
\centering
\includegraphics[width=0.60\linewidth]{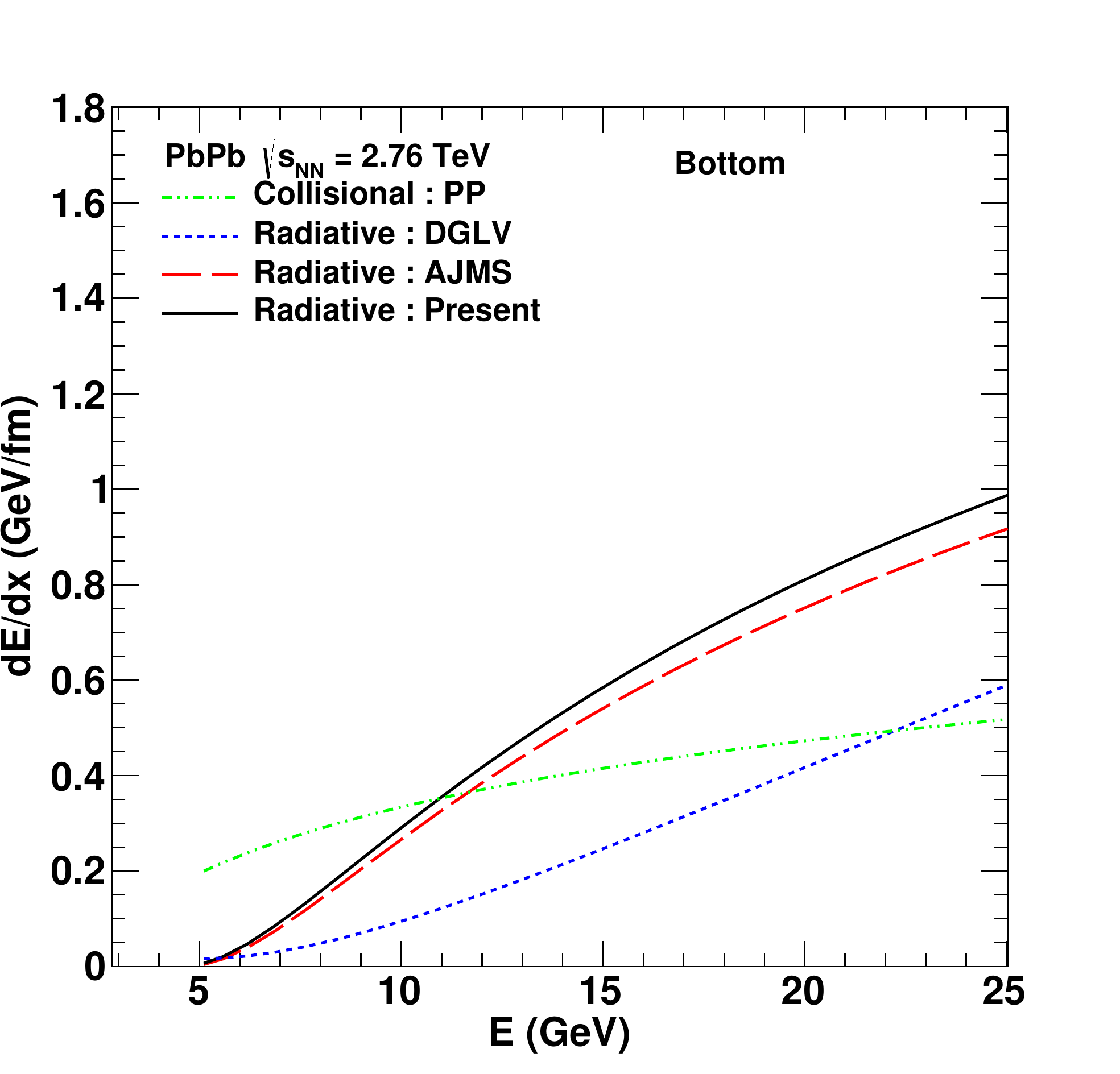} 
\caption{(color online): The energy loss $dE/dx$ as a function of energy
of bottom quark obtained using PP, DGLV, AJMS and corrected 
AJMS (Present) calculations in PbPb collision at $\sqrt{s_{NN}}$=2.76 TeV.}
\label{figure7bottom276energyloss}
\end{figure}

     Figure $\ref{figure4charm200energyloss}$ shows energy loss of charm quark 
as a function of energy of quark 
for AuAu collision at $\sqrt{s_{NN}}$=200 GeV using PP, DGLV, AJMS and corrected AJMS 
formalisms. Figure $\ref{figure5bottom200energyloss}$ is the same for bottom quark. 
  It can be seen that the collisional energy loss is similar in magnitude for 
charm and bottom quark. 
  The radiative energy loss calculated by AJMS is larger than that by DGLV. 
This difference is more pronounced for bottom quark. Radiative energy loss of 
bottom quark 
by AJMS starts dominating collisional energy loss at quark energy above 11 GeV whereas 
the DGLV energy loss remains below collisional energy loss upto 25 GeV of 
bottom quark energy. 
 Figure $\ref{figure6charm276energyloss}$ shows energy loss of charm quark 
as a function of energy of quark 
for PbPb collision at $\sqrt{s_{NN}}$=2.76 TeV using PP, DGLV, AJMS and corrected AJMS 
formalisms. Figure $\ref{figure7bottom276energyloss}$ is the same for bottom quark. 
 When we move from RHIC to LHC both the collisional as well as radiative energy loss increase.
  The radiative energy loss of charm quark calculated by AJMS and DGLV are similar in magnitudes
but the DGLV energy loss increases more steeply with quark energy.  
 For bottom quark, the AJMS energy loss is much larger than the DGLV energy loss.
Radiative energy loss of bottom quark 
by AJMS starts dominating collisional energy loss at quark energy above 10 GeV whereas 
the DGLV energy loss remains below collisional energy loss upto 22 GeV of 
bottom quark energy. 

\begin{figure}[htp]
\centering
\includegraphics[width=0.60\linewidth]{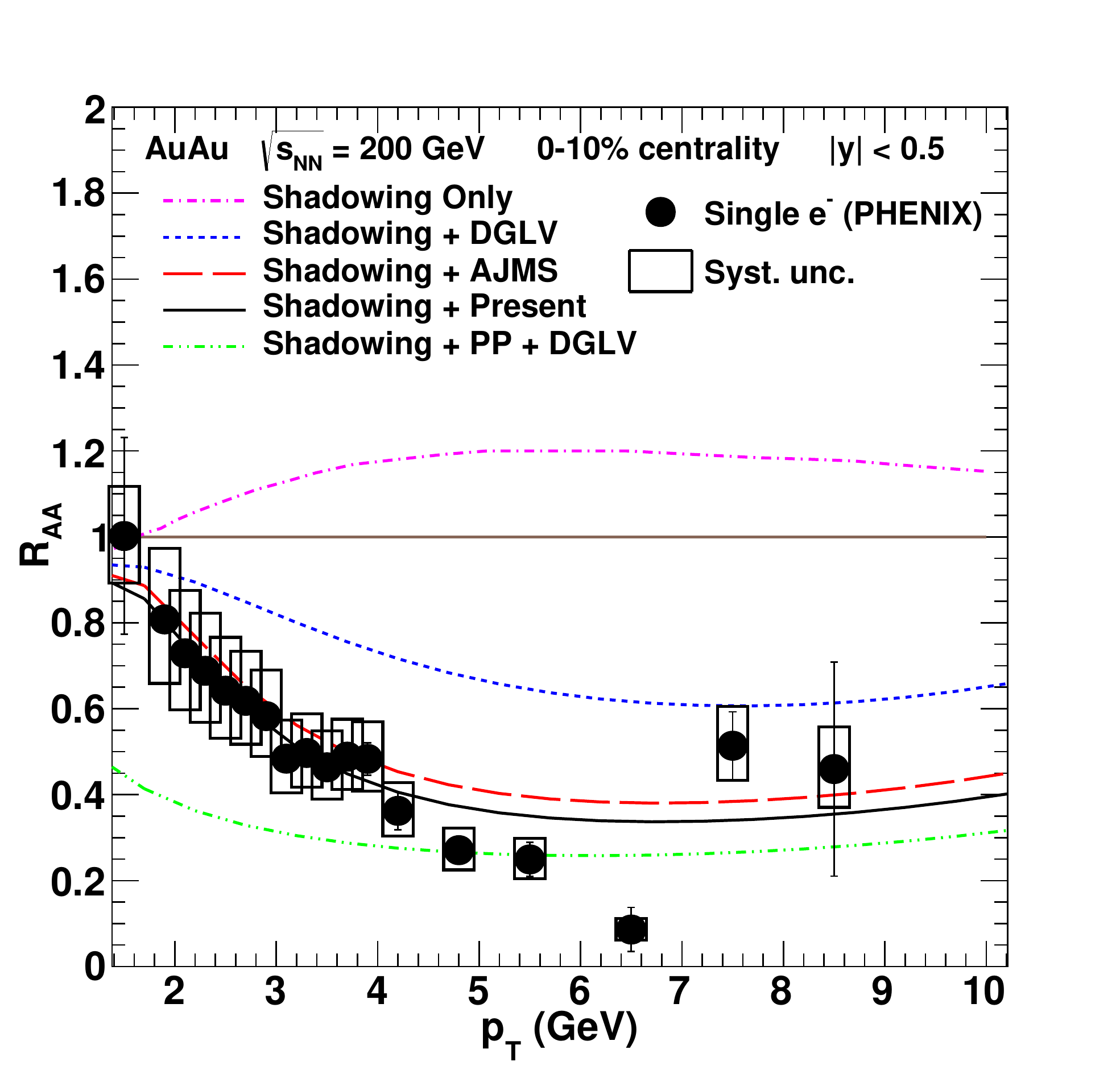} 
\caption{(color online): Nuclear modification factor $R_{AA}$ of single electrons 
from $D$ meson  as a function of transverse momentum $p_T$ obtained using 
energy loss (DGLV, AJMS, corrected AJMS (Present) and DGLV+PP calculations) and 
shadowing in AuAu collision at $\sqrt{s_{NN}}$=200 GeV.
The data is from PHENIX measurements of  heavy flavour (HF) electrons~\cite{adare}.}
\label{figure8phenix200raa}
\end{figure}

   Figure \ref{figure8phenix200raa} shows nuclear modification factor $R_{AA}$ of single 
electrons from $D$ meson  as a function of transverse momentum $p_T$ obtained using 
energy loss (DGLV, AJMS, corrected AJMS  and DGLV+PP calculations) and 
shadowing in AuAu collision at $\sqrt{s_{NN}}$=200 GeV.
The data is PHENIX measurements of  heavy flavour (HF) electrons~\cite{adare}. 
We observe that radiative energy loss by DGLV added to collisional energy loss by 
PP describes the PHENIX data at high $p_{T}$ range.
The radiative energy loss by AJMS and corrected AJMS reproduce the data without adding 
energy loss due to collisions.

\begin{figure}[htp]
\centering
\includegraphics[width=0.60\linewidth]{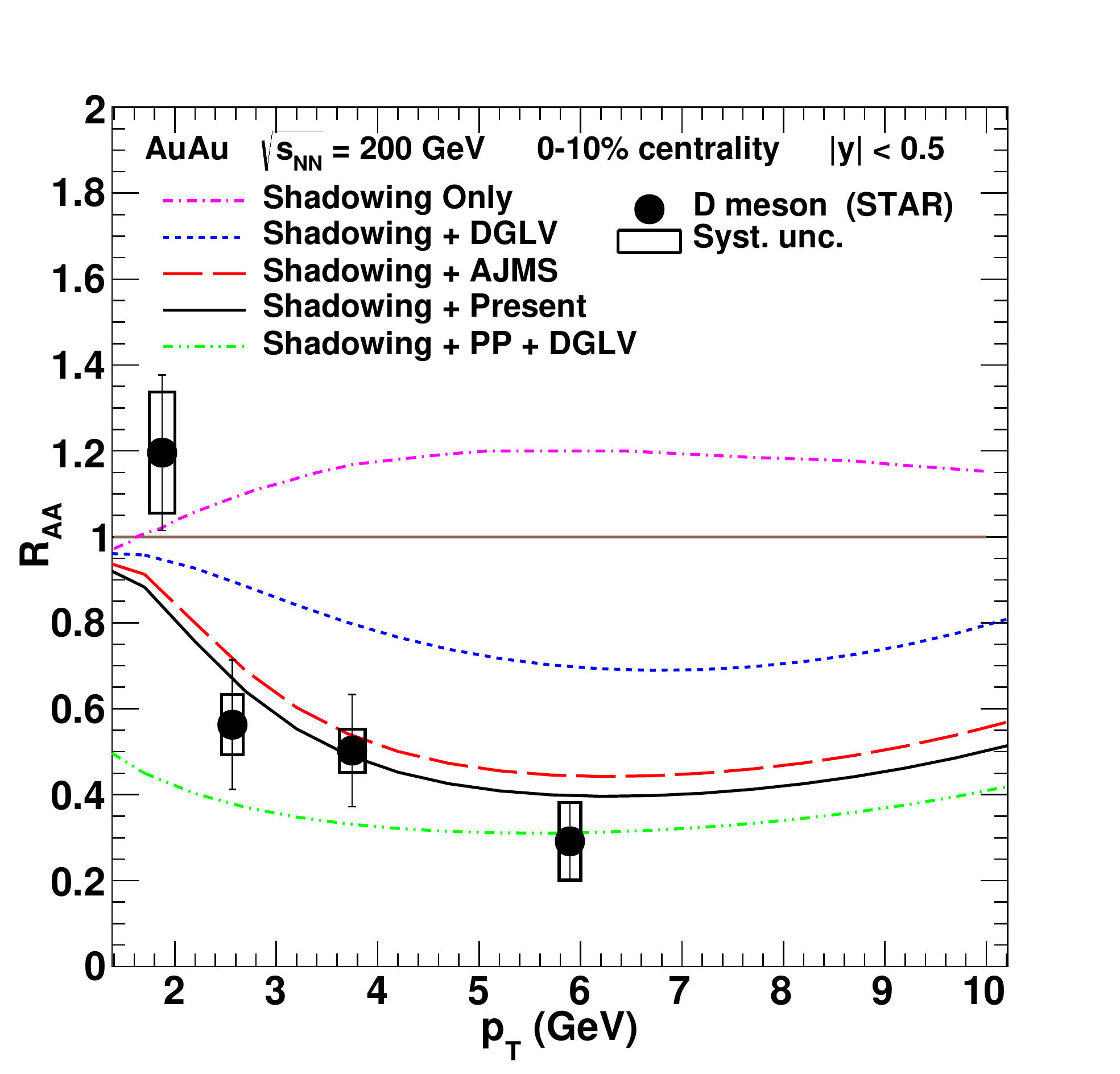} 
\caption{(color online): Nuclear modification factor $R_{AA}$ of $D$ meson  
as a function of transverse momentum $p_T$ obtained using 
energy loss (DGLV, AJMS, corrected AJMS (Present) and DGLV+PP calculations) and 
shadowing in AuAu collision at $\sqrt{s_{NN}}$=200 GeV. The data is from STAR 
measurements of $D$ mesons \cite{Adamczyk:2014uip}.}
\label{figure9star200raa}
\end{figure}

    Figure \ref{figure9star200raa} shows the nuclear modification factor $R_{AA}$ 
of $D$ meson as a function of transverse momentum $p_{T}$ obtained using energy loss 
energy loss (DGLV, AJMS, corrected AJMS  and DGLV+PP calculations) and 
shadowing in AuAu collision at $\sqrt{s_{NN}}$=200 GeV. The data is STAR 
measurements of $D^{0}$ mesons \cite{Adamczyk:2014uip}. We observe that the radiative 
energy loss by AJMS and corrected AJMS reproduce the data without adding energy loss 
due to collisions. The radiative energy loss by DGLV added to collisional energy loss by 
PP describes the STAR data at high $p_{T}$ range. 

\begin{figure}[htp]
\centering
\includegraphics[width=0.60\linewidth]{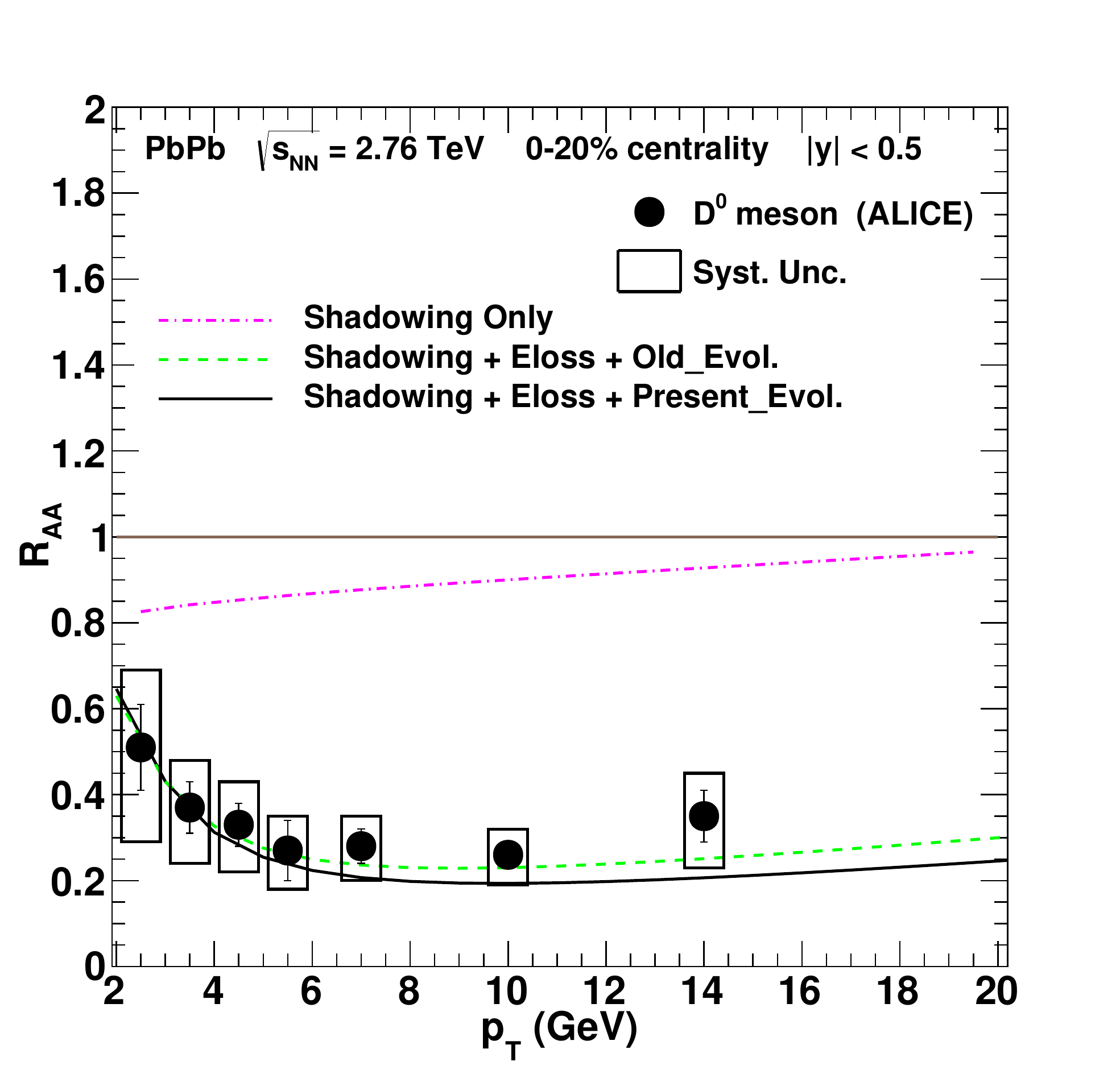} 
\caption{(color online): Nuclear modification factor $R_{AA}$ of $D^{0}$ mesons as a 
function of transverse momentum obtained using radiative energy loss (corrected AJMS 
(Present) calculations) calculated with old and new evolution models and shadowing in 
PbPb collision at $\sqrt{s_{NN}}$= 2.76 TeV. The data is from ALICE measurements 
of $D^{0}$ mesons~ \cite{alice:2012}.}
\label{figure10aliceraaoldandnewmodel}
\end{figure}

      Figure \ref{figure10aliceraaoldandnewmodel} shows the nuclear modification factor 
$R_{AA}$ of $D^{0}$ mesons as a function of transverse momentum obtained using radiative 
energy loss (corrected AJMS calculations) calculated with old and new evolution models 
and shadowing in PbPb collision at $\sqrt{s_{NN}}$= 2.76 TeV. The data is ALICE measurements 
of $D^{0}$ mesons~ \cite{alice:2012}
The value of $R_{AA}$ depends on energy loss model as well as evolution model
used to calculate the pathlength. 

\begin{figure}[htp]
\centering
\includegraphics[width=0.60\linewidth]{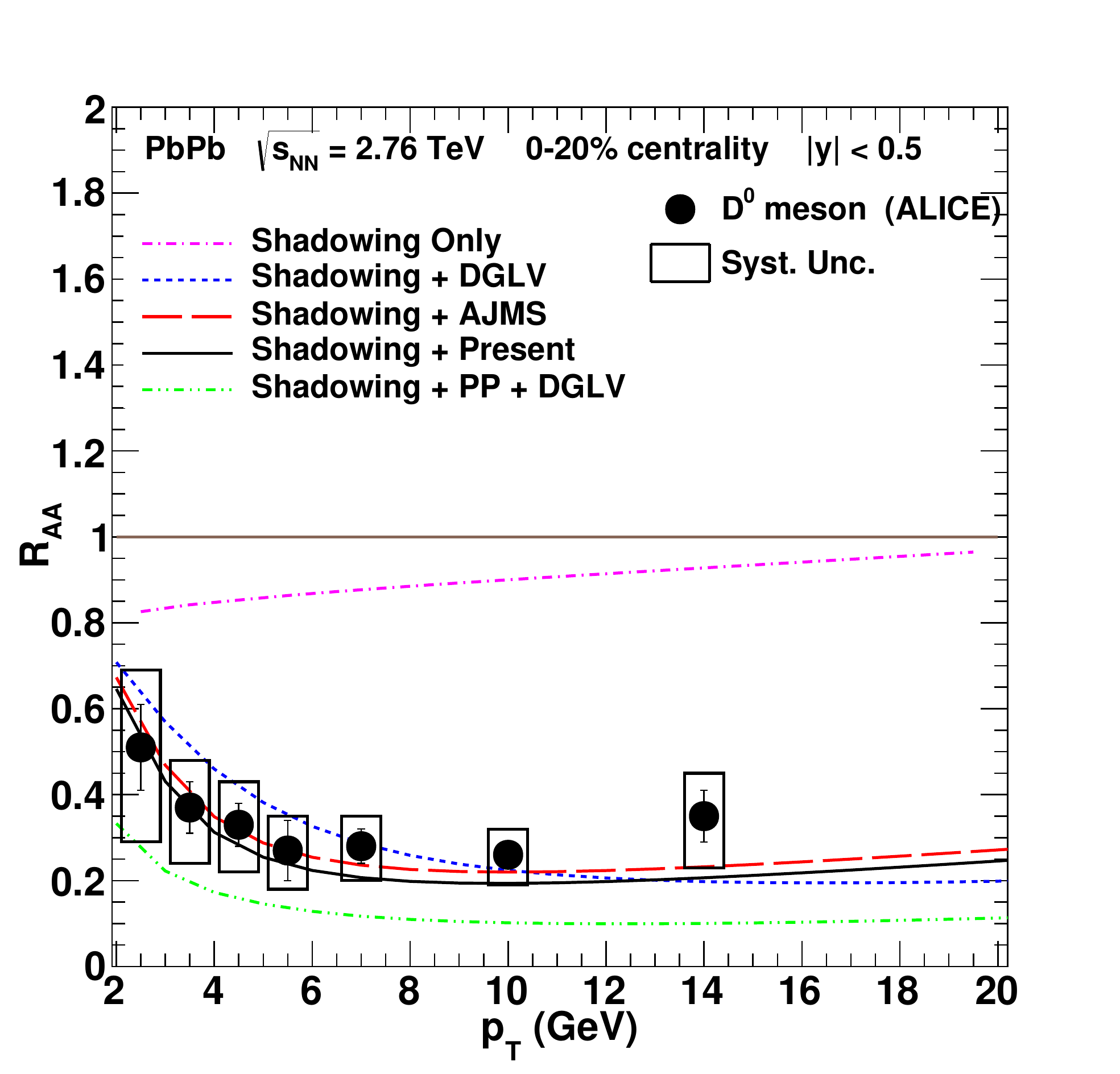} 
\caption{(color online): Nuclear modification factor $R_{AA}$ of $D^{0}$ mesons
as a function of transverse momentum $p_T$ obtained using energy 
loss (DGLV, AJMS, corrected AJMS (Present) and PP+DGLV calculations) and shadowing in 
PbPb collision at $\sqrt{s_{NN}}$= 2.76 TeV. The data is from ALICE measurements of $D^{0}$ 
mesons~ \cite{alice:2012}.}
\label{figure11alice276raa}
\end{figure}

      Figure \ref{figure11alice276raa} shows the nuclear modification factor $R_{AA}$ of 
$D^{0}$ mesons as a function of transverse momentum $p_T$ obtained using energy 
loss (DGLV, AJMS, corrected AJMS and PP+DGLV calculations) and shadowing in 
PbPb collision at $\sqrt{s_{NN}}$= 2.76 TeV. The data is ALICE measurements of $D^{0}$ 
mesons~ \cite{alice:2012}.  AJMS , corrected AJMS and DGLV calculations produce similar 
suppression in high $p_{T}$ range. When we add radiative and collisional energy 
loss (PP+DGLV) it overestimates the measured suppression of D meson.

\begin{figure}[htp]
\centering
\includegraphics[width=0.60\linewidth]{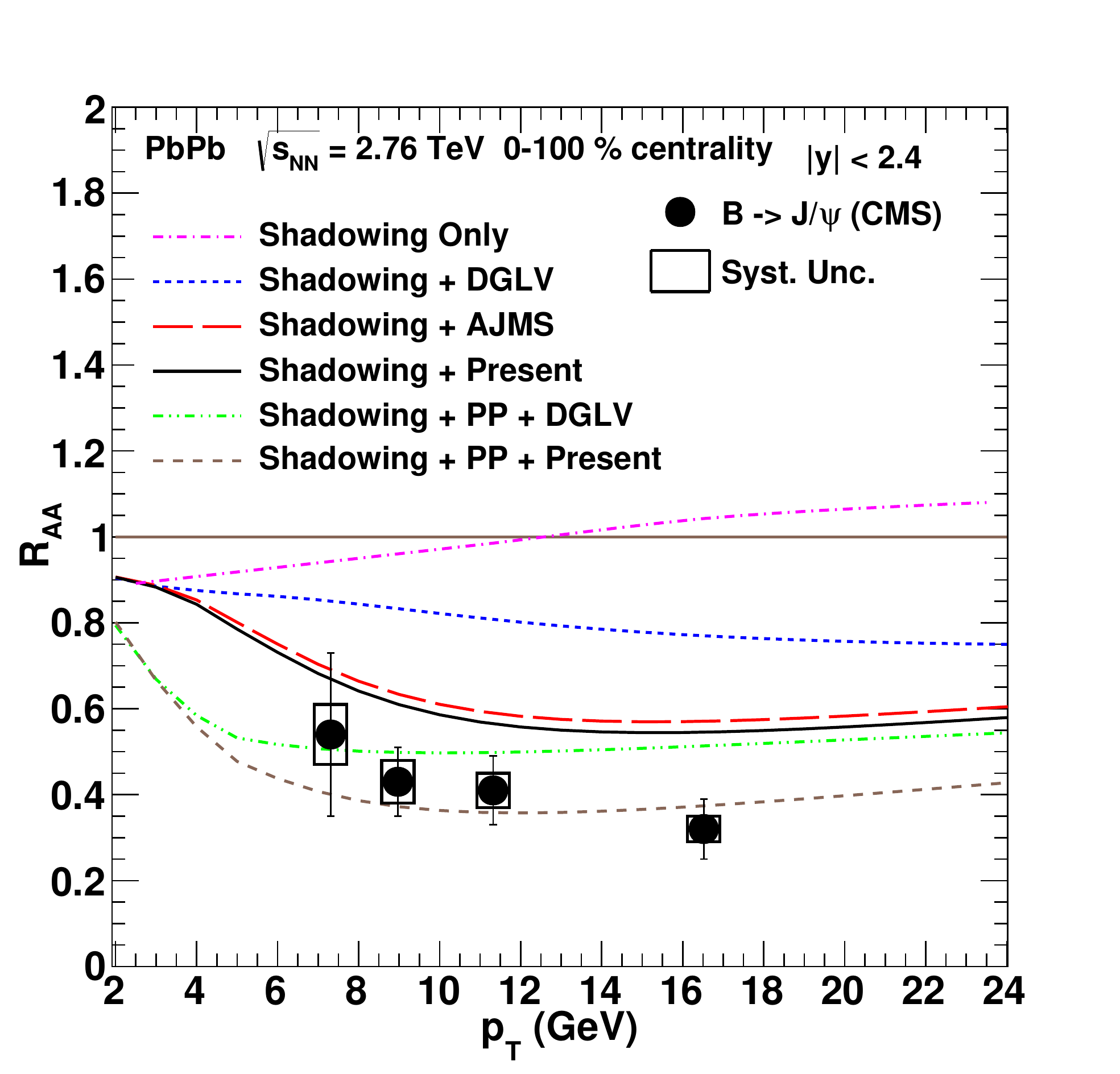} 
\caption{(color online): Nuclear modification factor $R_{AA}$ inclusive 
$J/\psi$ coming from $B$ mesons as a function of transverse momentum $p_T$ obtained 
using energy loss (DGLV, AJMS, corrected AJMS (Present) and PP+DGLV calculations) and 
shadowing in  PbPb collision at $\sqrt{s_{NN}}$= 2.76 TeV. The data is from CMS measurements 
of $J/\psi$ mesons from $B$ decays ~~\cite{cms}.}
\label{figure12cms276raa}
\end{figure}

       Figure \ref{figure12cms276raa}  shows the nuclear modification factor $R_{AA}$ 
inclusive $J/\psi$ coming from $B$ mesons as a function of transverse momentum 
$p_T$ obtained using energy loss (DGLV, AJMS, corrected AJMS and PP+DGLV 
calculations) and shadowing in  PbPb collision at $\sqrt{s_{NN}}$= 2.76 TeV. The data 
is CMS measurements of $J/\psi$ mesons from $B$ decays \cite{cms}.
We observe that sum of radiative energy loss (DGLV) and collisions energy loss (PP) 
underestimates the $B$ meson suppression. 
 The sum of radiative energy loss by corrected AJMS and collisions energy loss 
slightly overestimates the suppression. More accurate data in larger $p_T$ range would
help distinguish the models more clearly.

\begin{figure}[htp]
\centering
\includegraphics[width=0.60\linewidth]{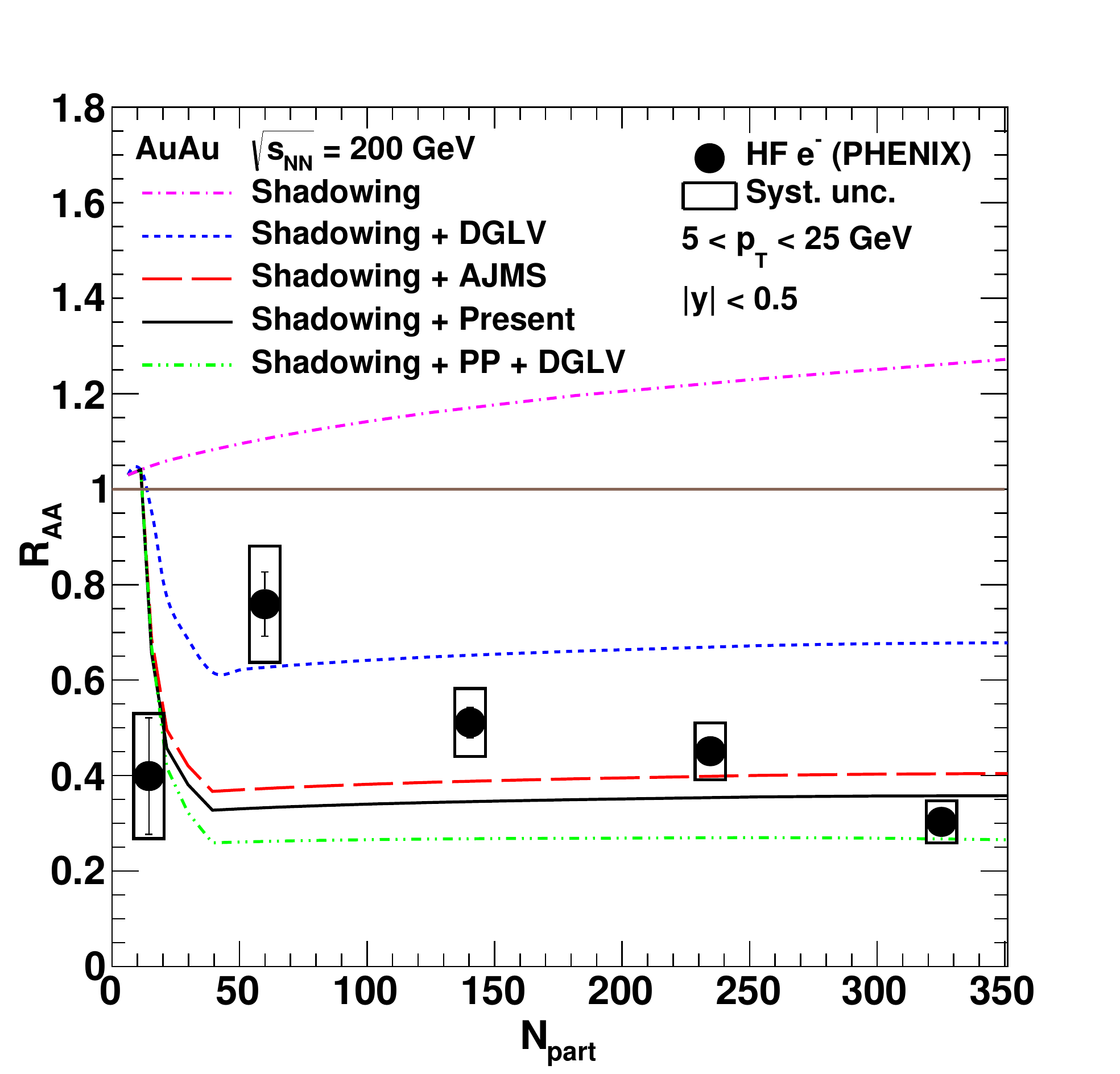} 
\caption{(color online): Nuclear modification factor $R_{AA}$ 
of single electrons from $D$ meson as a function of the  
number of participant $N_{part}$ 
obtained using energy loss (DGLV, AJMS, corrected AJMS (Present) and PP+DGLV 
calculations) and shadowing in AuAu collision  at $\sqrt{s_{NN}}$=200 GeV.
The data is from PHENIX measurements of heavy flavour (HF) electrons~\cite{adare}.}
\label{figure13phenixraanaprt200gev}
\end{figure}

 Figure~\ref{figure13phenixraanaprt200gev} shows 
 the nuclear modification factor $R_{AA}$ 
of single electrons from $D$ meson as a function of the  
number of participant $N_{part}$ 
obtained using energy loss (DGLV, AJMS, corrected AJMS (Present) and PP+DGLV 
calculations) and shadowing in AuAu collision  at $\sqrt{s_{NN}}$=200 GeV
compared with the PHENIX measurements of heavy flavour (HF) electrons~\cite{adare}.
  The radiative energy loss by DGLV added to collisional energy loss 
by PP slightly overestimates the suppression.
The radiative energy loss by corrected AJMS describes the data 
without adding energy loss due to collisions.

\begin{figure}[htp]
\centering
\includegraphics[width=0.60\linewidth]{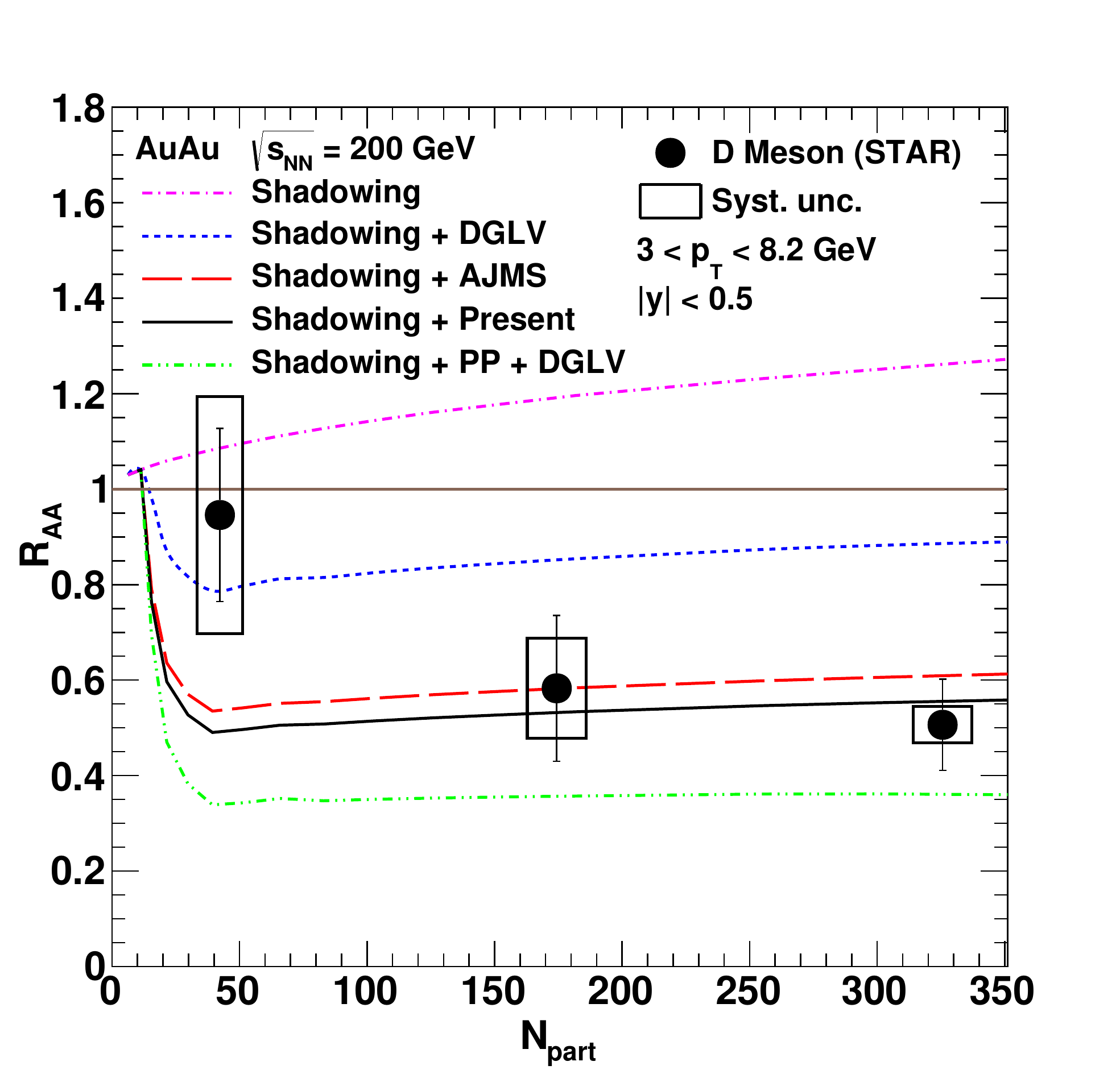}
\caption{(color online): Nuclear modification factor $R_{AA}$ 
of single electrons from $D$ meson as a function of the  
number of participant $N_{part}$ obtained using energy loss (DGLV, AJMS, 
corrected AJMS (Present) and PP+DGLV calculations) and shadowing in AuAu 
collision  at $\sqrt{s_{NN}}$=200 GeV. The data is from STAR measurements of 
$D$ mesons \cite{Adamczyk:2014uip}.}
\label{figure14starraanaprt200gev}
\end{figure}

    Figure~\ref{figure14starraanaprt200gev} shows the nuclear modification factor 
$R_{AA}$ of single electrons from $D$ meson as a function of the number of 
participant $N_{part}$  obtained using energy loss (DGLV, AJMS, corrected AJMS 
and PP+DGLV calculations) and shadowing in AuAu collision  at $\sqrt{s_{NN}}$ = 200 GeV
compared with the STAR measurements of $D$ mesons \cite{Adamczyk:2014uip}. We observe 
that the radiative energy loss by AJMS and corrected AJMS describe the data without 
adding energy loss due to collisions. The energy loss by DGLV does not describe the 
data.

\begin{figure}[htp]
\centering
\includegraphics[width=0.60\linewidth]{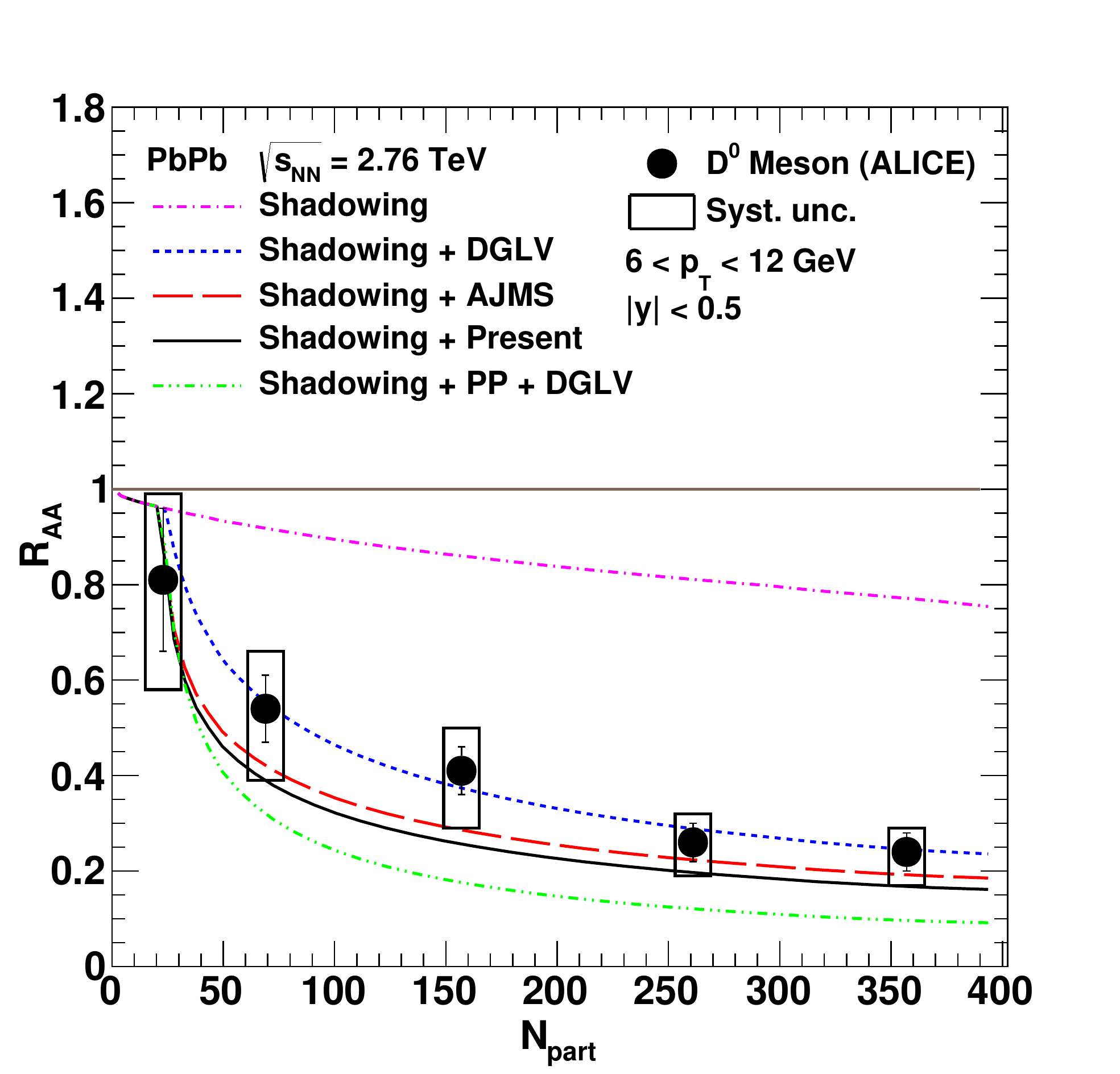}
\caption{(color online): Nuclear modification factor $R_{AA}$ of $D^{0}$ mesons 
as a function of the number of participant $N_{part}$ obtained using energy 
loss (DGLV, AJMS, corrected AJMS (Present) and PP+DGLV calculations) and 
shadowing in the PbPb collision  at  $\sqrt{s_{NN}}$=2.76 TeV.
The data is from ALICE measurements of $D^{0}$ mesons ~\cite{alice:2012}. }
\label{figure15alicedmesonraanaprt276tev}
\end{figure}

     Figure~\ref{figure15alicedmesonraanaprt276tev} shows the 
the nuclear modification factor $R_{AA}$ of $D^{0}$ mesons 
as a function of the number of participant $N_{part}$ obtained using energy 
loss (DGLV, AJMS, corrected AJMS (Present) and PP+DGLV calculations) and 
shadowing in the PbPb collision  at  $\sqrt{s_{NN}}$=2.76 TeV compared with
ALICE measurements of $D^{0}$ mesons ~\cite{alice:2012}.
  We observe that the radiative 
energy loss by DGLV, AJMS and corrected AJMS describe the ALICE data. 
The radiative energy loss by DGLV added to collisional energy loss by PP 
overestimates the $D^{0}$ suppression.
  
\begin{figure}[htp]
\centering
\includegraphics[width=0.60\linewidth]{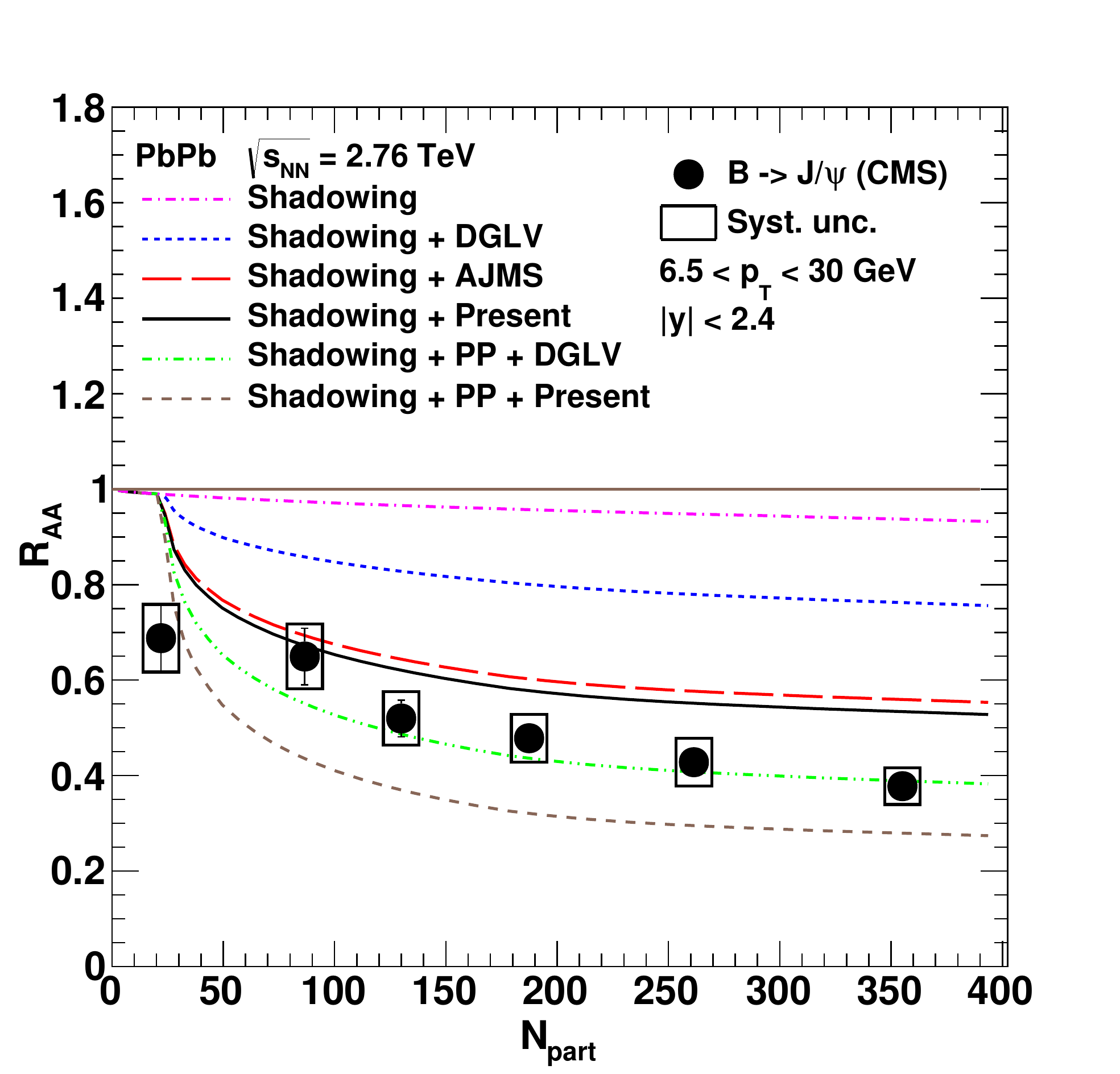}
\caption{(color online): Nuclear modification factor $R_{AA}$ of
inclusive $J\psi$ coming from $B$ mesons as a function of the  
number of participant $N_{part}$ obtained using energy 
loss (DGLV, AJMS, corrected AJMS (Present), PP+DGLV and PP+corrected AJMS calculations) 
and shadowing in the PbPb collision  at  $\sqrt{s_{NN}}$= 2.76 TeV.
The data is from CMS measurements of $J/\psi$ mesons from $B$ decays ~\cite{cms}.}
\label{figure16cmsbmesonraanaprt276tev}
\end{figure}

    Figure~\ref{figure16cmsbmesonraanaprt276tev} shows the 
nuclear modification factor $R_{AA}$ of
inclusive $J/\psi$ coming from $B$ mesons as a function of the  
number of participant $N_{part}$ obtained using energy 
loss (DGLV, AJMS, corrected AJMS (Present), PP+DGLV and PP+corrected AJMS calculations) 
and shadowing in the PbPb collision  at  $\sqrt{s_{NN}}$= 2.76 TeV compared with
the CMS measurements of $J/\psi$ mesons from $B$ decays \cite{cms}.
   We observe that the radiative energy loss by DGLV added to collisional energy loss 
by PP describes the CMS data very well. The radiative energy loss by 
corrected AJMS underestimates the suppression but with collisional energy loss 
added it overestimates the suppression. Both the models favour that 
B meson loose energy by radiation as well as collision processes.

\section{Conclusion}
  We study the energy loss of heavy quark (charm and bottom) due to elastic collisions and 
gluon radiation in hot/dense medium. Results of Radiative energy loss obtained from two 
different formalisms namely DGLV and AJMS have been compared. 
  The energy loss calculated by AJMS exceeds DGLV results.
The collisional energy loss has been calculated using Peigne and Peshier formalism and
is found to be similar in magnitude for charm and bottom quarks.
  The nuclear modification factors $R_{AA}$ including shadowing and 
energy loss are evaluated for $B$ and $D$ mesons and are compared with the 
measurements in PbPb collision at $\sqrt{s_{NN}}$ = 2.76 TeV and with the 
HF electrons and $D^{0}$ meson measurements in AuAu collision at 
$\sqrt{s_{NN}}$ = 200 GeV. 
  The radiative plus collisional energy loss (PP+DGLV) describes the 
RHIC HF electron suppression in high $p_{T}$ range. It also describes the LHC measurement
of $B$ meson suppression but overestimates the suppression of $D$ meson.
  The radiative energy loss from generalized dead cone approach describes the charm 
suppression at both RHIC as well as LHC energies without requiring collisional energy loss.
 Both collision as well as radiative energy loss are required to explain the $B$ meson
suppression at LHC. 
 Upcoming high luminosity PbPb collisions at $\sqrt{s_{NN}}$ = 5 TeV are 
expected to measure the heavy quarks in wider kinematic ranges 
which will provide much improved constrains for the processes of energy loss and models.

{\bf Acknowledgements} \\
 We are thankful to Umme Jamil, M. G. Mustafa , D. K. Shrivastava and V. Kumar for many
fruitful discussions.

{\bf Appendix A: DGLV Formalism} \\
 The average radiative energy loss of heavy quarks as
\begin{eqnarray}
 \frac{\Delta E}{L} &=& E~\frac{C_{F} \alpha_{s}}{\pi} 
\frac{1}{\lambda} \int^{1-\frac{M}{E+p}}_{\frac{m_g}{E+p}} dx \int^{\infty}_0 
\frac{4~ \mu^2_{g}~ q^3~ dq}{\left( \frac{4Ex}{L} \right)^2 + (q^2 + \beta^2)^2} 
\times (A \log B + C)~~,
\end{eqnarray}
where      
\begin{eqnarray} 
\beta^2 &=& m^2_g (1-x) + M^2 x^2 ,~~\lambda^{-1} = {\rho_g \sigma_{Qg} + 
\rho_q \sigma_{Qq} }~~, \\
\rho_{g} &=& 16~T^{3}~\frac{1.202}{\pi^{2}},~~~\rho_{q}=9~N_{f}~T^{3}~
\frac{1.202}{\pi^{2}},~~     \\
\sigma_{Qq} &=& \frac{9\pi\alpha_{s}^2}{2 \mu_{g}^2} \mbox{ and } \sigma_{Qg} = 
\frac{4}{9} \sigma_{Qq}\; \; .
\end{eqnarray}
\noindent
Here $C_{F}(=4/3)$ determines the coupling strength of gluon to the massive quark
with momentum $p$. $\rho_{g}$ and $\rho_{q}$ are the densities 
of gluons and quarks and $m_g = {\mu_{g}}/{\sqrt{2}}$ is the transverse gluon mass.

\noindent
The function $A$, $B$ and $C$ are given as follows
 \begin{eqnarray}
 A &=& \frac{2 \beta^2}{f_{\beta}^3} \left( \beta^2 + q^2 \right) ,
\\ 
 B &=& \frac{(\beta^2 + K) (\beta^2 Q^-_{\mu} + Q^+_{\mu}Q^+_{\mu}  + 
Q^+_{\mu} f_{\beta} )}{ 
\beta^2 \Big(\beta^2 (Q^-_{\mu} -K) - Q^-_{\mu}K + Q^+_{\mu}Q^+_{\mu} + f_{\beta}
f_{\mu}\Big) }, 
\\
 C &=& \frac{1}{2 q^2 f_{\beta}^2 f_{\mu}}  
\left[ \beta^2 \mu_{g}^2 (2 q^2 - \mu_{g}^2) + \beta^2 (\beta^2 - \mu_{g}^2) K \right.
\nonumber \\ 
&\;& + Q^+_{\mu} (\beta^4 - 2 q^2 Q^+_{\mu}) 
  +
   f_{\mu} 
\Big(\beta^2 (-\beta^2 - 3 q^2 +\mu_{g}^2) \nonumber\\
&\;& \left. + 2 q^2 Q^+_{\mu}\Big) + 3 \beta^2 q^2 Q^-_{k} \right].
 \end{eqnarray}
Here 
\begin{eqnarray}
 K&=& k_{max}^2 = 2p x(1-x), \\
Q^\pm_\mu &=& q^2 \pm \mu_{g}^2, ~~~
Q^\pm_k   = q^2\pm k_{max}^2, \\
f_\beta &=& f(\beta, Q^-_\mu,Q^+_\mu), ~~f_\mu =f(\mu_{g},Q^+_k,Q^-_k), \\
f(x,y,z) &=&\sqrt{x^4+2x^2 y+ z^2}.
\end{eqnarray}

{\bf Appendix B: Corrected AJMS} \\
The integration of Eq. (\ref{energyperunitdistance}) are obtained as follows.\\
The minimum values of $q^{2}_{\perp}$,~$\omega$ and $k^{2}_{\perp}$ are given 
by infra-red cut-off ~\cite{rumd,mustafa,dpal, xiang}
\begin{equation}
\label{minimumlimit}
q^{2}_{\perp}|_{min}~\approx~\omega^{2}_{min}~\approx~k^{2}_{\perp}|_{min}~
\approx~\mu^{2}_{g}.
\end{equation}

\noindent
The maximum value of $q^{2}_{\perp}|_{max}$ is calculated as~
\cite{dpal, xiang, pal} 
\begin{equation}
\label{maximumlimit}
q^{2}_{\perp}|_{max} = C~E~T~,
\end{equation}
where 
\begin{eqnarray}
C=\frac{3}{2}-\frac{M^{2}}{4~E~T}+\frac{M^{4}}{48~E^{2}~T^{2}~\beta_{0}}~
\log\Big[\frac{M^{2}+6~E~T~(1+\beta_{0})}{M^{2}+6~E~T~(1-\beta_{0})}\Big]
\end{eqnarray}
and 
\begin{eqnarray}
\beta_{0}=\sqrt{1-\frac{M^{2}}{E^{2}}}~.
\end{eqnarray}

\noindent
The maximum value of $\omega$ is obtained as ~\cite{wgp}
\begin{equation}
\label{omegamax}
\omega^{2}_{max}=<q^{2}_{\perp}>~.
\end{equation}

\noindent
The average of square of the transverse momentum $q_{\perp}$ is given in 
reference ~\cite{dpal,xiang} as 
\begin{equation}
\label{qavergae}
<q^{2}_{\perp}>=\frac{q^{2}_{\perp}|_{min}~q^{2}_{\perp}|_{max}}
{q^{2}_{\perp}|_{max}-q^{2}_{\perp}|_{min}}~
\log\Big[\frac{q^{2}_{\perp}|_{max}}{q^{2}_{\perp}|_{min}}\Big]~.
\end{equation}
Putting $q^{2}_{\perp}|_{min}$ from Eq. $(\ref{minimumlimit})$ and 
$q^{2}_{\perp}|_{max}$ from 
Eq. $(\ref{maximumlimit})$ in Eq. $(\ref{qavergae})$
\begin{equation}
\label{qsquareav}
<q^{2}_{\perp}> = \frac{\mu^{2}_{g}}{(1-\beta_{1})}~
\log\Big[\frac{1}{\beta_{1}} \Big],
\end{equation}
where $\beta_{1}=~\mu^{2}_{g}/(C~E~T)$.
Using the relation $\omega= k_{\perp}~\cosh\eta$,~ the finite cut on $\omega$ 
and $k_{\perp}$ leads to an inequality
\begin{equation}
\label{inequality}
\frac{\omega_{max}}{k_{\perp}|_{min}}~~  \textless ~~ \cosh\eta.
\end{equation}

\noindent
The integration limits of $\eta$ are calculated from Eq.~$(\ref{minimumlimit})$,~ 
$(\ref{omegamax})$ and $(\ref{inequality})$ as
\begin{equation}
\label{modeta}
|\eta| < \log\Bigg(\sqrt{\frac{<q^{2}_{\perp}>}{\mu^{2}_{g}}}+ 
\sqrt{\frac{<q^{2}_{\perp}>}{\mu^{2}_{g}}-1} \Bigg).
\end{equation}

\noindent
We can write it as  $|\eta|< \delta $, where
$\delta$ is obtained using equation  $(\ref{qsquareav})$ and $(\ref{modeta})$
\begin{equation}
\delta=\frac{1}{2}~\log\Bigg[\frac{1}{(1-\beta_{1})}~\log\Big(\frac{1}
{\beta_{1}}\Big)~\Bigg(1+\sqrt{1-\frac{(1-\beta_{1})}{\log(\frac{1}{\beta_{1}})}} 
\Bigg)^{2} \Bigg].
\end{equation}

\noindent
We can write the minimum and maximum value of $\eta$ as 
\begin{equation}
\eta_{min}=-\delta ,~~~~ \eta_{max}=\delta . 
\end{equation}

\noindent
Now we calculate the integrals in Eq. (\ref{energyperunitdistance}) 
which can be written as 
\begin{equation}
\label{energyajms}
\frac{dE}{dx}=24~\alpha^{3}_{s}~\rho_{QGP}~I_{1}~I_{2}~I_{3}~~.
\end{equation}
The first integration $I_{1}$ is calculated as 
\begin{equation}
\label{qsquarecal}
I_{1} = \int^{C~E~T}_{\mu^{2}_{g}}~\frac{1}{(q^{2}_{\perp})^{2}}~
dq^{2}_{\perp}  = \frac{1}{\mu^{2}_{g}}~\Big(1-\beta_{1} \Big).
\end{equation}

\noindent
The second integration $I_{2}$ is calculated as 
\begin{equation}
\label{omegacal}
I_{2} = \int^{\omega_{max}}_{\omega_{\min}}~d\omega =
 \mu_{g}~\Bigg(\sqrt{\frac{1}{(1-\beta_{1})}~
\log\Big(\frac{1}{\beta_{1}} \Big)}~~~-1 \Bigg).
\end{equation}

\noindent
The third integration $I_{3}$ is calculated as 
\begin{eqnarray}
I_{3} &=& \int^{\eta_{max}}_{\eta_{min}}~\mathcal D~d\eta = \int^{\delta}_{-\delta}~
\frac{1}{\Big(1+\frac{M^{2}}{s}~e^{2\eta}\Big)^{2}}~d\eta~~,\\
 &=& \frac{1}{2}~\Bigg[\frac{1}{1+\frac{M^2}{s}~e^{2\eta}}+
\log\Bigg(\frac{\frac{M^2}{s}~e^{2\eta}}{1+\frac{M^2}{s}~e^{2\eta}}
 \Bigg)\Bigg]^{\delta}_{-\delta}~~,\\
   &=& \frac{1}{2}~\Bigg[\log(e^{4\delta}) + \log\Bigg
(\frac{1+\frac{M^2}{s}~e^{-2\delta}}{1+\frac{M^2}{s}~e^{2\delta}}\Bigg)-
\Bigg(\frac{\frac{M^{2}}{s}~(e^{2\delta}-e^{-2\delta})}
{1+\frac{M^2}{s}~(e^{2\delta}+e^{-2\delta})+\frac{M^4}{s^2}} \Bigg)\Bigg].
\end{eqnarray}

\noindent
The integration $I_{3}$ is denoted as $\mathcal F(\delta)$ given by
\begin{equation}
\mathcal F(\delta)=2\delta-\frac{1}{2}~\log\Bigg(
\frac{1+\frac{M^2}{s}~e^{2\delta}}{1+\frac{M^2}{s}~e^{-2\delta}}\Bigg)-
\Bigg(\frac{\frac{M^2}{s}~\sinh(2\delta)}
{1+2~\frac{M^2}{s}\cosh(2\delta)+\frac{M^4}{s^{2}}}\Bigg)~~.
\end{equation}

{\bf References}


\end{document}